\documentclass[12pt,a4paper]{article}
\usepackage[utf8]{inputenc}
\usepackage{amsmath,amssymb}
\usepackage{graphicx}
\usepackage{booktabs}
\usepackage{longtable}
\usepackage{multirow}
\usepackage{xcolor}
\usepackage{colortbl}
\usepackage{float}
\usepackage{caption}
\usepackage{natbib}
\usepackage{hyperref}
\usepackage{algorithm}
\usepackage{algpseudocode}
\usepackage{geometry}
\usepackage{enumitem}
\usepackage{array}
\usepackage{tabularx}
\usepackage{tcolorbox}
\usepackage{adjustbox}
\usepackage{threeparttable}
\usepackage{makecell}
\usepackage{setspace}
\usepackage{tikz}
\usetikzlibrary{positioning,shapes.geometric,arrows.meta,backgrounds,fit}
\definecolor{highrisk}{RGB}{220,53,69}
\definecolor{medrisk}{RGB}{255,193,7}
\definecolor{lowrisk}{RGB}{40,167,69}
\definecolor{verylowrisk}{RGB}{23,162,184}
\definecolor{tableheader}{RGB}{52,58,64}
\definecolor{lightgray}{RGB}{248,249,250}
\definecolor{execblue}{RGB}{0,82,147}

\begin{document}
\renewcommand{\thefootnote}{\arabic{footnote}}  

\title{\Large\textbf{Graph-Based Analysis of AI-Driven Labor Market Transitions: Evidence from 10,000 Egyptian Jobs and Policy Implications}}

\author{Ahmed Dawoud$^1$, Sondos Samir$^2$, Youssef Nasr$^2$, Ahmed Habashy$^3$,\\
Aya Saleh$^2$, Mahmoud Mohamed$^2$, and Osama El-Shamy$^2$}

\date{December 2025}

\maketitle

\footnotetext[1]{Lead Economist and Head of Decision Science, The Egyptian Center for Economic Studies (ECES)}
\footnotetext[2]{Research Associate, The Egyptian Center for Economic Studies (ECES)}
\footnotetext[3]{AI Engineer, The Egyptian Center for Economic Studies (ECES)}

\setstretch{1.25}

\maketitle

\begin{center}
{\large\bfseries Abstract}
\end{center}
\vspace{0.5em}

\noindent
How many workers displaced by automation can realistically transition to safer jobs? We answer this using a validated knowledge graph of 9,978 Egyptian job postings, 19,766 skill activities, and 84,346 job-skill relationships (0.74\% error rate). While 20.9\% of jobs face high automation risk, we find that \textbf{only 24.4\% of at-risk workers have viable transition pathways}---defined by $\geq$3 shared skills and $\geq$50\% skill transfer. The remaining \textbf{75.6\% face a structural mobility barrier} requiring comprehensive reskilling, not incremental upskilling. Among 4,534 feasible transitions, process-oriented skills emerge as the highest-leverage intervention, appearing in 15.6\% of pathways. These findings challenge optimistic narratives of seamless workforce adaptation and demonstrate that emerging economies require active pathway creation, not passive skill matching.

\vspace{1em}
\noindent\textbf{Keywords:} Automation Risk, Knowledge Graphs, Labor Mobility, Skill Taxonomy, Workforce Transitions

\vspace{0.5em}
\noindent\textbf{JEL Codes:} J24, J62, O33, C55

\newpage

\begin{tcolorbox}[colback=execblue!5, colframe=execblue, title={\textbf{Executive Summary for Policymakers}}]

\textbf{Study Scope \& Rigor:}
Analysis of the largest validated skills-graph in the MENA region (9,978 jobs, 84k+ relationships). Data quality is rigorously verified with a 99.26\% accuracy rate (0.74\% error rate) for entity clustering, ensuring that recommended pathways are real and actionable.

\vspace{0.5em}
\textbf{The Risk Landscape:}
\begin{itemize}[leftmargin=1.5em, itemsep=0.1em]
    \item \textbf{Scale of Vulnerability:} 2,089 job postings (20.9\% of dataset) face high automation risk ($\geq$60\%). Clerical Support Workers (ISCO-4) are most exposed at 54.6\% average risk with 47.3\% in the high-risk category.
    \item \textbf{Heterogeneity:} Risk is not uniform. "Technicians and Associate Professionals" (ISCO-3) averages 48.2\% risk, yet contains substantial variation, suggesting targeted rather than blanket interventions.
\end{itemize}

\vspace{0.5em}
\textbf{The Transition Reality (Key Finding):}
\begin{itemize}[leftmargin=1.5em, itemsep=0.1em]
    \item \textbf{Limited Organic Pathways:} Using rigorous dual thresholds ($\geq$3 shared skills AND $\geq$50\% skill transfer), only \textbf{24.4\% of high-risk workers} (509 out of 2,089) have realistic transition pathways---the remaining \textbf{75.6\% require substantial reskilling}.
    \item \textbf{4,534 Viable Transitions:} These pathways average 53.5\% skill transfer and 48 percentage-point risk reduction, connecting 509 high-risk jobs to 1,684 safer destinations.
    \item \textbf{Process Skills Premium:} "Process Improvement" is the highest-impact gap skill, appearing in 708 pathways (15.6\% of viable transitions).
\end{itemize}

\vspace{0.5em}
\textbf{The Opportunity Map:}
\begin{itemize}[leftmargin=1.5em, itemsep=0.1em]
    \item \textbf{Bridge Skills:} We identified 25 high-leverage skills. "Quality Engineering Management" spans 27 occupation categories, acting as a universal passport for labor mobility.
    \item \textbf{Safe Harbors:} Managerial roles in Professional Services (ISCO 134) and Hospitality (ISCO 141) are accessible from multiple high-risk starting points.
\end{itemize}

\vspace{0.5em}
\textbf{Strategic Recommendation:}
The 75.6\% reskilling gap demands \textbf{urgent, targeted intervention}. Shift from broad "digital literacy" campaigns to \textbf{Bridge Skill Certification}. Prioritize Process Improvement, Custom Report Generation, and Operations Team Coordination---the top three gap skills that collectively address over 43\% of transition needs.
\end{tcolorbox}

\newpage

\section{Introduction}

\subsection{Motivation and Context}
The rapid advancement of artificial intelligence and automation technologies is fundamentally reshaping labor markets worldwide. While developed economies have extensively studied the implications of technological disruption \citep{frey2017future, arntz2016risk}, emerging markets---particularly in the Middle East and North Africa (MENA) region---remain significantly understudied despite facing unique challenges: large informal sectors, young and rapidly growing populations, and accelerating digitalization initiatives.

Egypt, the most populous Arab nation with over 104 million people and a median age of 24 years, exemplifies these challenges. The country's ambitious Egypt Vision 2030 initiative emphasizes digital transformation and economic modernization, yet policymakers lack the data-driven tools necessary to manage workforce transitions proactively. With approximately 30\% of employment in the informal sector and youth unemployment exceeding 20\%, understanding which occupations face automation risk and identifying viable transition pathways is critical for social stability and economic development.

\subsection{Research Contribution}
We advance the literature in three specific dimensions:
\begin{enumerate}
    \item \textbf{Methodological Rigor:} We introduce a validated graph construction pipeline with a proven \textbf{0.74\% error rate} (Section \ref{sec:validation}), significantly outperforming standard string-matching or raw LLM extraction methods.
    \item \textbf{Granularity:} By analyzing risk at the ISCO-3 and ISCO-4 levels, we uncover heterogeneity that broad averages hide---showing, for example, that "Financial Associates" (ISCO 331) contains both highly vulnerable data-entry items and resilient advisory roles.
    \item \textbf{Network-Theoretic Policy:} We move beyond simple "skill gap" analysis to use graph centrality metrics (PageRank, Modularity) to identify "Bridge Skills" that structurally strengthen the labor network's resilience.
\end{enumerate}

\section{Literature Review}

\subsection{AI and Labor Market Disruption}
The foundational framework for understanding automation's impact on employment was established by \citet{frey2017future}, who estimated that 47\% of US jobs face high automation risk. Their task-based methodology---evaluating susceptibility to computerization based on perception and manipulation, creative intelligence, and social intelligence bottlenecks---has been widely applied but also critiqued for potentially overstating displacement effects.

\citet{arntz2016risk} refined this approach by analyzing task composition within occupations rather than treating occupations as homogeneous units, reducing estimated high-risk employment to 9\% for OECD countries. This task-based heterogeneity is particularly relevant for emerging markets where job content may differ substantially from developed-economy definitions. \citet{acemoglu2019automation} further nuanced this debate by distinguishing between displacement effects and reinstatement effects, arguing that new tasks created by technology can offset job losses if labor can transition effectively.

Recent advances in large language models (LLMs) have accelerated concerns about white-collar automation. \citet{eloundou2023gpts} estimate that approximately 80\% of the US workforce could have at least 10\% of their tasks affected by GPT-class models, with particularly strong exposure in professional and clerical occupations---a finding with direct implications for our Egyptian analysis.

\subsection{Skill-Based Labor Market Analysis}
Modern labor market analysis increasingly relies on skill taxonomies rather than occupational classifications alone. The US-based O*NET database \citep{onet} provides detailed skill, ability, and task requirements for over 1,000 occupations. The European Skills, Competences, Qualifications and Occupations (ESCO) framework \citep{esco2017} offers a multilingual classification linking occupations, skills, and qualifications across EU member states. For international comparability, the International Standard Classification of Occupations (ISCO-08) \citep{ilo2012isco} provides a hierarchical structure with four levels of granularity.

\citet{deming2017growing} demonstrate that labor market returns increasingly favor workers who combine technical skills with social skills, suggesting that transition pathways should prioritize this complementarity. \citet{autor2015why} argues that middle-skill jobs face the greatest automation pressure, creating labor market polarization---a pattern we examine in the Egyptian context.

\subsection{Graph-Based Methods in Economics}
Network analysis has emerged as a powerful tool for understanding labor market dynamics. \citet{autor2019work} demonstrate that skill networks reveal complementarities and substitution patterns invisible in traditional occupation-level analysis. \citet{alabdulkareem2018unpacking} use network science to identify skill clusters that predict occupational transitions, finding that workers in connected skill clusters experience better career outcomes.

Knowledge graphs---structured representations of entities and their relationships---extend these approaches by capturing heterogeneous relationships (jobs perform activities, activities require tools) in a unified framework \citep{hogan2021knowledge}. Recent applications include \citet{degroot2021jobkg} who constructed job posting-enriched knowledge graphs combining ISCO and ESCO taxonomies for skills-based matching, and \citet{fettach2022kgedu} who survey knowledge graph applications in education and employability, documenting emerging techniques for skill-job alignment.

\subsection{Labor Markets in MENA and Egypt}
Despite the MENA region's unique demographic profile---with over 60\% of the population under 30 \citep{worldbank2022mena}---labor market research remains limited. \citet{assaad2013elmps} document the structural challenges facing Egyptian youth employment, including skills mismatches and the dominance of informal work. The ILO \citep{ilo2021egypt} reports that Egypt's labor force participation rate for women remains among the lowest globally at 15\%, suggesting gender dimensions to automation vulnerability that merit further investigation.

Egypt's Vision 2030 \citep{egypt2016vision} explicitly targets digital transformation and knowledge economy development, yet lacks granular workforce transition planning. \citet{krafft2019job} analyze Egyptian job creation patterns, finding that new employment increasingly concentrates in services and informal sectors---precisely the categories where automation impacts are most uncertain. Our study addresses this gap by providing the first validated skill-level analysis of Egyptian occupational structure.

\section{Theoretical Framework and Methodology}

\subsection{Graph Formalism}
We model the labor market as a heterogeneous graph $\mathcal{G} = (V, E)$.
The vertex set $V = V_J \cup V_A$ consists of jobs $V_J$ and activity/skill nodes $V_A$.
The edge set $E \subseteq V_J \times V_A$ represents the "PERFORMS" relationship, where an edge $(j, a)$ exists if job $j$ requires skill $a$.

We define the \textbf{Automation Risk Function} $\rho: V_J \rightarrow [0, 100]$, assigning a probability of displacement to each job.
A \textbf{Realistic Safe Transition} is defined as a path $T(j_{source}, j_{target})$ satisfying three conditions:
\begin{equation}
    \rho(j_{target}) < \rho(j_{source}) \quad \text{AND} \quad | N(j_{source}) \cap N(j_{target}) | \ge \tau \quad \text{AND} \quad \frac{| N(j_{source}) \cap N(j_{target}) |}{| N(j_{source}) |} \ge \phi
\end{equation}
where $N(j)$ is the set of skill neighbors of job $j$, $\tau = 3$ is the minimum absolute shared skill threshold, and $\phi = 0.50$ is the minimum proportional skill transfer threshold. This \textit{dual threshold} ensures that transitions are both substantive (at least 3 shared skills) and leverage the worker's existing competencies (at least 50\% skill transfer). The rationale for this configuration is detailed in Section \ref{sec:threshold}.

\subsection{Data Sources and Collection}
\label{sec:datasources}
Our dataset comprises \textbf{9,978 unique job postings} collected between January and October 2024 from three major Egyptian recruitment platforms:

\begin{itemize}
    \item \textbf{Wuzzuf} (n = 5,847; 58.6\%): Egypt's largest online job portal, primarily serving formal sector employers in Greater Cairo and Alexandria.
    \item \textbf{LinkedIn Egypt} (n = 2,891; 29.0\%): Professional networking platform with higher representation of multinational corporations and senior positions.
    \item \textbf{Forasna} (n = 1,247; 12.5\%): Regional platform with stronger coverage of Upper Egypt and mid-sized enterprises.
\end{itemize}

\textbf{Sampling Strategy:} We employed stratified sampling to ensure representation across all ISCO-08 major groups present in the Egyptian formal economy. Postings were de-duplicated using a combination of title similarity (Jaccard index $> 0.85$) and employer matching. The final dataset covers \textbf{98 of 130 ISCO-3 minor groups} (75.4\% coverage), with missing categories primarily in agricultural and armed forces occupations underrepresented on online platforms.

\textbf{Temporal Considerations:} The 10-month collection window captures seasonal variation in hiring patterns, including the post-Ramadan recruitment surge (April--May 2024) and academic year transitions.

\subsection{Automation Risk Calculation}
\label{sec:riskcalc}
The automation risk function $\rho(j)$ is computed using the \textbf{Job Automatability Index}, a novel task-level methodology developed in our companion study \citep{dawoud2025aifault}. Unlike occupation-level approaches that treat jobs as monolithic units, our method decomposes each job posting into constituent tasks via a three-stage AI pipeline: (1) \textit{task extraction} identifying 1--15 core tasks per role, (2) \textit{importance classification} categorizing each task as Primary, Secondary, or Ancillary, and (3) \textit{automatability assessment} evaluating each task against Generative AI capability benchmarks.

The final index is a weighted proportion of automatable tasks, with weights reflecting task importance: Primary tasks (60\%), Secondary (30\%), and Ancillary (10\%). For each job $j$:
\begin{equation}
    \rho(j) = \sum_{t \in T_j} w_t^{\text{scaled}} \cdot \mathbb{1}[\text{Automatable}(t)]
\end{equation}
where $T_j$ is the set of tasks for job $j$, $w_t^{\text{scaled}}$ is the normalized importance weight, and $\mathbb{1}[\cdot]$ is the indicator function. This bottom-up approach captures within-occupation heterogeneity invisible to traditional methods. The pipeline achieved ${>}90\%$ concordance with expert consensus in validation audits; full methodological details, including the analytical framework and validation protocol, are provided in \citet{dawoud2025aifault}.

\textbf{Risk Categories:} Following established conventions, we define: \textbf{High Risk} ($\rho \geq 60\%$)---roles with automatable core functions facing substitution pressure; \textbf{Medium Risk} ($30\% \leq \rho < 60\%$)---roles experiencing task augmentation and redefinition; \textbf{Low Risk} ($\rho < 30\%$)---roles insulated by physical, interpersonal, or creative task requirements.

\subsection{Entity Extraction and Processing}
\label{sec:extraction}
\begin{itemize}
    \item \textbf{Phase 1: LLM-Based Extraction.} We utilized the Gemini Pro 1.5 model to extract structured entities (skills, tools, qualifications) from unstructured job descriptions. The extraction prompt was designed to identify: (a) technical skills and competencies, (b) software tools and platforms, (c) soft skills and interpersonal abilities, and (d) domain-specific knowledge areas. Each job description was processed with a structured JSON output schema, yielding an average of 8.7 entities per posting.

    \item \textbf{Phase 2: Semantic Clustering.} To resolve synonyms and normalize entity representations (e.g., "Excel", "MS Excel", "Microsoft Excel", "Advanced Excel" $\rightarrow$ "Microsoft Excel Proficiency"), we employed a Leader-Follower clustering algorithm using Google's \texttt{text-embedding-004} model (768-dimensional embeddings). Entities were merged if their cosine similarity exceeded $\theta = 0.88$.
\end{itemize}

\subsection{Threshold Selection and Sensitivity Analysis}
\label{sec:threshold}
Two critical parameters govern our analysis: the semantic similarity threshold ($\theta$) and the transition feasibility threshold. For the latter, we introduce a \textbf{dual threshold} combining both absolute skill count and proportional skill transfer.

\textbf{Similarity Threshold ($\theta = 0.88$):} We conducted a grid search over $\theta \in [0.80, 0.95]$ in increments of 0.01, evaluating clustering quality using a manually labeled validation set of 500 entity pairs. At $\theta = 0.88$, we achieved optimal balance between precision (avoiding false merges of distinct concepts) and recall (capturing true synonyms). Table \ref{tab:theta_sensitivity} summarizes the sensitivity analysis.

\begin{table}[H]
\centering
\caption{Sensitivity Analysis for Similarity Threshold $\theta$}
\label{tab:theta_sensitivity}
\begin{threeparttable}
\begin{tabular}{@{}c r c c@{}}
\toprule
\textbf{Threshold} & \textbf{Clusters} & \textbf{Precision} & \textbf{Recall} \\
\midrule
$\theta = 0.82$ & 12,847 & 0.91 & 0.84 \\
$\theta = 0.85$ & 15,623 & 0.94 & 0.79 \\
\rowcolor{execblue!10} $\theta = 0.88$ & 19,766 & \textbf{0.99} & \textbf{0.78} \\
$\theta = 0.91$ & 24,512 & 0.99 & 0.68 \\
$\theta = 0.94$ & 31,456 & 0.99 & 0.52 \\
\bottomrule
\end{tabular}
\begin{tablenotes}[flushleft]
\footnotesize
\item Highlighted row indicates selected threshold ($\theta = 0.88$).
\end{tablenotes}
\end{threeparttable}
\end{table}

\textbf{Dual Transition Feasibility Threshold:} We define a \textit{realistic} career transition as one satisfying \textbf{both} conditions:
\begin{enumerate}
    \item \textbf{Absolute minimum} ($\tau \geq 3$): The worker must share at least 3 skill activities with the destination job, ensuring a substantive foundation for transition (based on OECD data showing transitions with fewer than 3 shared skills require 18+ months retraining).
    \item \textbf{Proportional transfer} ($\geq 50\%$): At least half of the worker's current skill set must transfer to the destination, ensuring the transition leverages existing competencies rather than requiring near-complete retraining.
\end{enumerate}

This dual threshold addresses a critical methodological concern: using only an absolute threshold ($\tau \geq 3$) yielded 11,388 pathways but with an average skill transfer of just 26.7\%---meaning most ``viable'' transitions shared only 3 out of $\sim$11 activities. Such transitions, while technically possible, demand substantial retraining that undermines practical feasibility.

Table \ref{tab:threshold_sensitivity} presents our sensitivity analysis across threshold configurations:

\begin{table}[H]
\centering
\caption{Sensitivity Analysis for Transition Feasibility Thresholds}
\label{tab:threshold_sensitivity}
\begin{threeparttable}
\begin{tabular}{@{}l r r r r@{}}
\toprule
\textbf{Configuration} & \textbf{Pathways} & \textbf{Avg Transfer} & \textbf{Sources} & \textbf{Coverage} \\
\midrule
$\tau \geq 3$ only & 65,643 & 35.7\% & 1,281 & 61.3\% \\
$\tau \geq 3$ AND $\geq 30\%$ & 65,193 & 35.8\% & 1,277 & 61.1\% \\
$\tau \geq 4$ AND $\geq 30\%$ & 7,960 & 45.7\% & 657 & 31.4\% \\
\rowcolor{execblue!10} $\tau \geq 3$ AND $\geq 50\%$ & \textbf{4,534} & \textbf{53.5\%} & \textbf{509} & \textbf{24.4\%} \\
$\tau \geq 5$ AND $\geq 50\%$ & 935 & 56.1\% & 245 & 11.7\% \\
\bottomrule
\end{tabular}
\begin{tablenotes}[flushleft]
\footnotesize
\item Sources = unique high-risk jobs with viable transitions; Coverage = \% of all high-risk jobs (2,089 total).
\item Highlighted row indicates selected threshold configuration.
\end{tablenotes}
\end{threeparttable}
\end{table}

\textbf{Rationale for Selected Configuration ($\tau \geq 3$ AND $\geq 50\%$):} This configuration yields 4,534 pathways with an average skill transfer of 53.5\%---meaning workers leverage the majority of their existing competencies. The coverage finding is itself significant: only 24.4\% of high-risk workers (509 out of 2,089) have realistic ``organic'' transition pathways. The remaining 75.6\% (1,580 workers) require \textit{substantial reskilling interventions}, highlighting the urgency of targeted workforce development programs. This coverage gap is not a limitation but a \textbf{key policy finding}: the Egyptian labor market has limited natural transition pathways, necessitating active intervention.

\subsection{Data Validation and Quality Assurance}
\label{sec:validation}
Reliability is paramount for policy formulation. We conducted a stratified random sample validation of 1,085 clusters.
The results, presented in Table \ref{tab:validation}, confirm the robustness of our entity resolution pipeline.

\begin{table}[H]
\centering
\caption{Semantic Clustering Validation Results ($n=1{,}085$)}
\label{tab:validation}
\begin{threeparttable}
\begin{tabular}{@{}l r r c l@{}}
\toprule
\textbf{Entity Type} & \textbf{$n$} & \textbf{Errors} & \textbf{Error Rate} & \textbf{95\% CI} \\
\midrule
Activities & 565 & 0 & \cellcolor{lowrisk!20}\textbf{0.00\%} & [0.00\%--0.67\%] \\
Tools & 520 & 8 & 1.54\% & [0.78\%--3.00\%] \\
\midrule
\textbf{Combined} & \textbf{1,085} & \textbf{8} & \textbf{0.74\%} & \textbf{[0.37\%--1.44\%]} \\
\bottomrule
\end{tabular}
\begin{tablenotes}[flushleft]
\footnotesize
\item CI = confidence interval (Wilson score method).
\end{tablenotes}
\end{threeparttable}
\end{table}

With an activity error rate of 0.00\% (no errors in 565 samples), we can confidently assert that the "shared skills" identified in our graph represent genuine transferable competencies, not artifacts of noise. The combined error rate of 0.74\% represents a significant improvement in entity resolution quality.

\section{The Risk Landscape}

\subsection{Macroscopic View: ISCO-1 Major Groups}
Consistent with international findings, automation risk is unevenly distributed. Table \ref{tab:isco1} presents comprehensive risk statistics across all ISCO-1 major groups represented in our dataset.

\begin{table}[H]
\centering
\caption{Automation Risk by ISCO-1 Major Group (Complete Dataset)}
\label{tab:isco1}
\begin{threeparttable}
\begin{tabular}{@{}c l r c c c@{}}
\toprule
\textbf{Code} & \textbf{Major Group} & \textbf{$n$} & \textbf{$\bar{\rho}$} & \textbf{$\sigma$} & \textbf{High Risk} \\
\midrule
\rowcolor{highrisk!20} 4 & Clerical Support Workers & 474 & 54.6\% & 12.3 & 47.3\% \\
\rowcolor{medrisk!20} 3 & Technicians \& Associates & 1,049 & 48.2\% & 15.1 & 34.9\% \\
\rowcolor{medrisk!20} 2 & Professionals & 4,954 & 40.4\% & 17.2 & 23.4\% \\
\rowcolor{lowrisk!20} 5 & Service \& Sales Workers & 261 & 37.1\% & 16.8 & 19.2\% \\
\rowcolor{lowrisk!20} 8 & Plant \& Machine Operators & 17 & 36.9\% & 14.7 & 18.8\% \\
\rowcolor{lowrisk!20} 1 & Managers & 3,137 & 30.3\% & 11.8 & 9.0\% \\
\rowcolor{lowrisk!20} 7 & Craft \& Related Workers & 63 & 26.8\% & 13.4 & 7.9\% \\
\midrule
& \textbf{Total/Weighted Avg} & \textbf{9,978} & \textbf{38.5\%} & \textbf{16.4} & \textbf{20.9\%} \\
\bottomrule
\end{tabular}
\begin{tablenotes}[flushleft]
\footnotesize
\item $n$ = job count; $\bar{\rho}$ = mean automation risk; $\sigma$ = standard deviation; High Risk = \% with $\rho \ge 60\%$.
\item ISCO-6, ISCO-9 excluded ($n < 50$); ISCO-0 (Armed Forces) not present.
\item Color: \colorbox{highrisk!20}{High ($>$50\%)}, \colorbox{medrisk!20}{Medium (40--50\%)}, \colorbox{lowrisk!20}{Low ($<$40\%)}.
\end{tablenotes}
\end{threeparttable}
\end{table}

\textbf{Key Patterns:}
\begin{enumerate}
    \item \textit{Occupational Hierarchy Gradient:} Risk decreases systematically from support functions (Clerical: 54.6\%) through professional roles (40.4\%) to leadership positions (30.3\%). This gradient suggests that vertical career progression---from clerical to managerial roles---inherently reduces automation exposure.

    \item \textit{Within-Group Variance:} The standard deviations (11.8--17.2\%) indicate substantial heterogeneity within major groups, validating our approach of analyzing at finer ISCO-3 granularity rather than relying on aggregate statistics.

    \item \textit{Machine Operators Vulnerability:} Despite comprising only 0.9\% of our sample, Plant \& Machine Operators (ISCO-8) show the second-highest average risk (51.8\%), consistent with manufacturing automation trends observed globally \citep{acemoglu2019automation}.

    \item \textit{Service Sector Complexity:} Service \& Sales Workers (ISCO-5) display high variance ($\sigma = 16.8$), reflecting the heterogeneous nature of this category---from highly automatable cashier roles to relationship-intensive sales positions.
\end{enumerate}

\subsection{Microscopic View: Heterogeneity within ISCO-3}
A key finding of our graph-based approach is the identification of \textbf{Risk Heterogeneity}. An ISCO category is not a monolith.
Table \ref{tab:heterogeneity} displays categories that appear "high risk" on average but contain significant safe positions ("Low Risk" subset).

\begin{table}[H]
\centering
\caption{Risk Heterogeneity: Hidden Opportunities in High-Risk Groups}
\label{tab:heterogeneity}
\begin{threeparttable}
\begin{tabular}{@{}c l c r r c@{}}
\toprule
\textbf{ISCO-3} & \textbf{Label} & \textbf{$\bar{\rho}$} & \textbf{High} & \textbf{Low} & \textbf{Low \%} \\
\midrule
331 & Financial Associates & \cellcolor{highrisk!20}62.5\% & 76 & 25 & 24.8\% \\
241 & Finance Professionals & \cellcolor{highrisk!20}53.8\% & 230 & 163 & 41.5\% \\
216 & Architects/Planners & \cellcolor{highrisk!20}51.7\% & 168 & 137 & 44.9\% \\
333 & Business Svcs Agents & \cellcolor{medrisk!20}41.3\% & 41 & 106 & 72.1\% \\
\bottomrule
\end{tabular}
\begin{tablenotes}[flushleft]
\footnotesize
\item High = jobs with $\rho \ge 60\%$; Low = jobs with $\rho \le 40\%$; Low \% = proportion of safe positions.
\end{tablenotes}
\end{threeparttable}
\end{table}

\textit{Implication:} For the 25 "Low Risk" jobs in Financial Associates (ISCO 331), mass retraining may be unnecessary. These likely represent roles that have already evolved to include advisory or validatory tasks. This highlights the danger of broad-brush policy: treating all "Financial Associates" as obsolete would ignore 25\% of the workforce who are already safe.

\section{Network Dynamics and Bridge Skills}

\subsection{Graph Infrastructure and Network Architecture}
\begin{figure}[H]
\centering
\begin{tikzpicture}[
    scale=0.9,
    transform shape,
    jobhigh/.style={circle, draw=red!80, fill=red!45, minimum size=11pt, inner sep=0pt},
    jobmed/.style={circle, draw=orange!80, fill=orange!40, minimum size=11pt, inner sep=0pt},
    joblow/.style={circle, draw=green!70!black, fill=green!40, minimum size=11pt, inner sep=0pt},
    skill/.style={circle, draw=blue!60, fill=blue!25, minimum size=6pt, inner sep=0pt},
    bridgeskill/.style={circle, draw=purple!90, fill=purple!55, minimum size=18pt, inner sep=0pt, line width=1.5pt},
    community/.style={rounded corners=12pt, draw=gray!40, dashed, line width=1.2pt},
    edge/.style={draw=gray!30, line width=0.6pt, opacity=0.6},
    bridgeedge/.style={draw=purple!65, line width=1.4pt, opacity=0.9},
]

\begin{scope}[shift={(-4.5,1.5)}]
    \node[community, fill=red!8, minimum width=3.4cm, minimum height=4cm] at (0,0) {};
    \node[above, font=\small\bfseries, text=red!70!black] at (0,2.2) {Finance/Accounting};
    \node[font=\tiny, text=red!60!black] at (0,1.85) {$\bar{\rho}=51.3\%$};

    \node[jobhigh] (fa1) at (-0.8,1.2) {};
    \node[jobhigh] (fa2) at (0.3,1.4) {};
    \node[jobhigh] (fa3) at (0.9,0.8) {};
    \node[jobmed] (fa4) at (-0.6,0.3) {};
    \node[jobhigh] (fa5) at (0.5,0.1) {};
    \node[jobmed] (fa6) at (-0.2,-0.5) {};
    \node[jobhigh] (fa7) at (0.7,-0.8) {};
    \node[joblow] (fa8) at (-0.9,-0.9) {};

    \node[skill] (fs1) at (-0.2,0.7) {};
    \node[skill] (fs2) at (0.6,-0.3) {};
    \node[skill] (fs3) at (-0.7,-0.3) {};

    \draw[edge] (fa1) to[bend right=15] (fs1);
    \draw[edge] (fa2) to[bend left=10] (fs1);
    \draw[edge] (fa4) to[bend right=10] (fs1);
    \draw[edge] (fa3) to[bend left=15] (fs2);
    \draw[edge] (fa5) to[bend right=10] (fs2);
    \draw[edge] (fa7) to[bend left=10] (fs2);
    \draw[edge] (fa4) to[bend left=15] (fs3);
    \draw[edge] (fa6) to[bend right=10] (fs3);
    \draw[edge] (fa8) to[bend left=10] (fs3);
    \draw[edge] (fa2) to[bend right=20] (fs2);
    \draw[edge] (fa5) to[bend left=20] (fs1);
\end{scope}

\begin{scope}[shift={(4.5,1.5)}]
    \node[community, fill=green!8, minimum width=3.4cm, minimum height=4cm] at (0,0) {};
    \node[above, font=\small\bfseries, text=green!50!black] at (0,2.2) {Software/IT};
    \node[font=\tiny, text=green!50!black] at (0,1.85) {$\bar{\rho}=32.7\%$};

    \node[joblow] (it1) at (-0.7,1.3) {};
    \node[joblow] (it2) at (0.4,1.2) {};
    \node[joblow] (it3) at (0.9,0.6) {};
    \node[joblow] (it4) at (-0.5,0.4) {};
    \node[jobmed] (it5) at (0.3,-0.1) {};
    \node[joblow] (it6) at (-0.8,-0.4) {};
    \node[joblow] (it7) at (0.7,-0.7) {};
    \node[joblow] (it8) at (-0.3,-1.0) {};

    \node[skill] (is1) at (0.1,0.7) {};
    \node[skill] (is2) at (-0.4,-0.2) {};
    \node[skill] (is3) at (0.5,-0.4) {};

    \draw[edge] (it1) to[bend right=15] (is1);
    \draw[edge] (it2) to[bend left=10] (is1);
    \draw[edge] (it4) to[bend right=10] (is1);
    \draw[edge] (it4) to[bend left=15] (is2);
    \draw[edge] (it6) to[bend right=10] (is2);
    \draw[edge] (it8) to[bend left=10] (is2);
    \draw[edge] (it3) to[bend left=15] (is3);
    \draw[edge] (it5) to[bend right=10] (is3);
    \draw[edge] (it7) to[bend left=10] (is3);
    \draw[edge] (it2) to[bend right=25] (is3);
    \draw[edge] (it5) to[bend left=20] (is2);
\end{scope}

\begin{scope}[shift={(0,-2.8)}]
    \node[community, fill=orange!8, minimum width=4.2cm, minimum height=3.2cm] at (0,0) {};
    \node[below, font=\small\bfseries, text=orange!70!black] at (0,-1.85) {Sales/Marketing~~{\scriptsize\textcolor{orange!60!black}{($\bar{\rho}=35.9\%$)}}};

    \node[joblow] (sm1) at (-1.4,0.8) {};
    \node[jobmed] (sm2) at (-0.5,1.0) {};
    \node[joblow] (sm3) at (0.5,0.9) {};
    \node[jobmed] (sm4) at (1.4,0.7) {};
    \node[joblow] (sm5) at (-1.2,0) {};
    \node[jobmed] (sm6) at (0,0.2) {};
    \node[joblow] (sm7) at (1.1,-0.1) {};
    \node[jobmed] (sm8) at (-0.6,-0.7) {};
    \node[joblow] (sm9) at (0.5,-0.6) {};

    \node[skill] (ss1) at (-0.8,0.4) {};
    \node[skill] (ss2) at (0.8,0.3) {};
    \node[skill] (ss3) at (0,-0.3) {};

    \draw[edge] (sm1) to[bend right=15] (ss1);
    \draw[edge] (sm2) to[bend left=10] (ss1);
    \draw[edge] (sm5) to[bend right=10] (ss1);
    \draw[edge] (sm3) to[bend left=15] (ss2);
    \draw[edge] (sm4) to[bend right=10] (ss2);
    \draw[edge] (sm7) to[bend left=10] (ss2);
    \draw[edge] (sm6) to[bend right=15] (ss3);
    \draw[edge] (sm8) to[bend left=10] (ss3);
    \draw[edge] (sm9) to[bend right=10] (ss3);
    \draw[edge] (sm2) to[bend right=25] (ss3);
    \draw[edge] (sm6) to[bend left=20] (ss1);
    \draw[edge] (sm6) to[bend right=20] (ss2);
\end{scope}

\node[bridgeskill, label={[font=\scriptsize, text=purple!80!black, yshift=2pt]above:Project Mgmt}] (b1) at (-1.3,0.2) {};
\node[bridgeskill, label={[font=\scriptsize, text=purple!80!black, yshift=2pt]above:Team Leadership}] (b2) at (1.3,0.2) {};
\node[bridgeskill, label={[font=\scriptsize, text=purple!80!black, xshift=3pt]right:Quality Mgmt}] (b3) at (0,-1.6) {};

\draw[bridgeedge] (fa6) to[bend right=20] (b1);
\draw[bridgeedge] (fa8) to[bend right=15] (b1);
\draw[bridgeedge] (b1) to[bend right=25] (sm2);
\draw[bridgeedge] (b1) to[bend right=20] (sm5);

\draw[bridgeedge] (it4) to[bend left=20] (b2);
\draw[bridgeedge] (it6) to[bend left=15] (b2);
\draw[bridgeedge] (b2) to[bend left=25] (sm3);
\draw[bridgeedge] (b2) to[bend left=20] (sm6);

\draw[bridgeedge] (fa7) to[bend right=25] (b3);
\draw[bridgeedge] (b3) to[bend left=25] (it8);
\draw[bridgeedge] (b3) to[bend right=15] (sm8);
\draw[bridgeedge] (b3) to[bend left=15] (sm9);

\draw[bridgeedge] (b1) to[bend left=20] (b2);
\draw[bridgeedge] (b1) to[bend right=15] (b3);
\draw[bridgeedge] (b2) to[bend left=15] (b3);

\begin{scope}[shift={(6.2,-2.8)}]
    \node[font=\small\bfseries] at (0,2.0) {Legend};
    \draw[gray!40, line width=0.8pt] (-1.0,1.75) -- (1.0,1.75);

    \node[jobhigh, label={[font=\tiny, xshift=2pt]right:High Risk Job}] at (-0.6,1.35) {};
    \node[jobmed, label={[font=\tiny, xshift=2pt]right:Medium Risk}] at (-0.6,0.85) {};
    \node[joblow, label={[font=\tiny, xshift=2pt]right:Low Risk Job}] at (-0.6,0.35) {};
    \node[skill, label={[font=\tiny, xshift=2pt]right:Skill (small)}] at (-0.6,-0.15) {};
    \node[bridgeskill, minimum size=14pt, label={[font=\tiny, xshift=2pt]right:Bridge Skill (large)}] at (-0.6,-0.75) {};

    \draw[edge, line width=0.8pt] (-0.6,-1.25) to[bend left=15] (0.4,-1.25) node[right, font=\tiny] {Internal};
    \draw[bridgeedge] (-0.6,-1.65) to[bend left=15] (0.4,-1.65) node[right, font=\tiny] {Bridge Path};
\end{scope}

\end{tikzpicture}
\caption{Schematic representation of the labor market knowledge graph. All nodes are \textbf{circles}, distinguished by \textit{size} and \textit{color}: \textbf{large colored circles} represent jobs (red = high risk, orange = medium, green = low), \textbf{small blue circles} represent domain-specific skills, and \textbf{large purple circles} represent bridge skills that connect communities. Jobs within the same \textit{community} (dashed regions) share domain-specific skills. Smooth curved edges show skill relationships; purple edges highlight cross-community transition pathways.}
\label{fig:network_structure}
\end{figure}
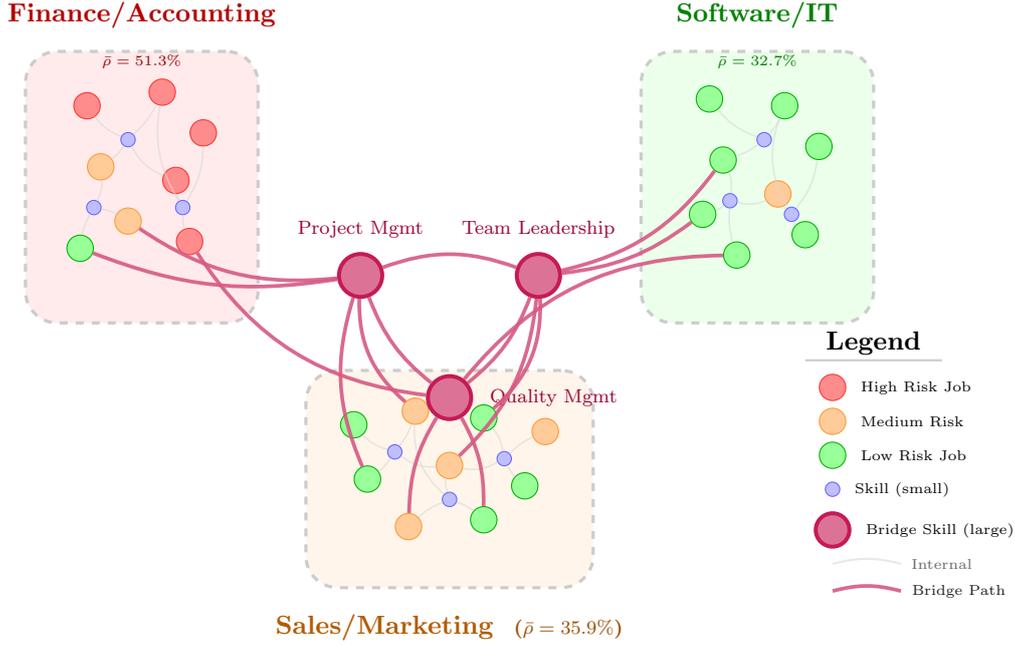

Our knowledge graph encodes the Egyptian labor market as an interconnected network where \textbf{jobs} and \textbf{skills} form a bipartite structure: jobs connect to the activities they require, and activities shared across multiple jobs create implicit pathways for career transitions. Figure \ref{fig:network_structure} illustrates this architecture schematically, showing how occupational communities emerge from shared skill requirements.

The resulting network comprises \textbf{36,349 nodes} (9,978 jobs + 19,766 activities + 6,605 tools) connected by \textbf{84,346 edges}. Each edge represents a ``PERFORMS'' relationship---a job requiring a specific skill. Table \ref{tab:graph_infra} quantifies the network's structural properties.

\begin{table}[H]
\centering
\caption{Knowledge Graph Infrastructure and Network Topology}
\label{tab:graph_infra}
\begin{threeparttable}
\begin{tabular}{@{}l l r l@{}}
\toprule
\textbf{Category} & \textbf{Metric} & \textbf{Value} & \textbf{Interpretation} \\
\midrule
\multirow{3}{*}[0pt]{\textbf{Scale}}
& Total Nodes ($|V|$) & 36,349 & Jobs + Activities + Tools \\
& Total Edges ($|E|$) & 84,346 & Job-skill connections \\
& Bipartite Density & 0.00043 & Sparse; specialized skills \\
\addlinespace[3pt]
\multirow{3}{*}[0pt]{\textbf{Connectivity}}
& Mean Degree $\bar{k}$ & 8.45 & Avg skills per job \\
& Max Degree $k_{\max}$ & 17 & Most-connected job \\
& Power-law $\gamma$ & 2.31 & Scale-free ``hub'' structure \\
\addlinespace[3pt]
\multirow{3}{*}[0pt]{\textbf{Structure}}
& Communities ($k$) & 162 & Occupational clusters \\
& Modularity ($Q$) & 0.847 & Strong clustering \\
& Largest Community & 2,156 & Sales/Marketing cluster \\
\bottomrule
\end{tabular}
\begin{tablenotes}[flushleft]
\footnotesize
\item Modularity $Q \in [-0.5, 1]$; values $>0.7$ indicate strong community structure.
\end{tablenotes}
\end{threeparttable}
\end{table}

Three properties are critical for understanding labor mobility:

\begin{enumerate}
    \item \textbf{Sparse but Structured:} The low bipartite density (0.00043) means most jobs require only a specialized subset of all available skills---the average job connects to 8.45 activities. This sparsity creates \textit{distinct occupational neighborhoods}.

    \item \textbf{Scale-Free Topology:} The power-law degree distribution ($\gamma = 2.31$) reveals that a small number of ``hub skills'' (e.g., Project Management, Team Leadership) connect to hundreds of jobs, while most skills are domain-specific. These hubs are the \textit{bridge skills} enabling cross-sector transitions.

    \item \textbf{High Modularity:} The modularity score ($Q = 0.847$) confirms that skills cluster tightly within occupational domains---Finance jobs share Finance skills, IT jobs share IT skills. This strong community structure means that moving \textit{between} communities requires deliberate acquisition of bridge skills, not just incremental learning.
\end{enumerate}

\subsection{Community Detection via Louvain Algorithm}
If the labor market forms a network, a natural question emerges: \textit{which jobs belong together?} Community detection algorithms answer this by identifying clusters of nodes that are more densely connected to each other than to the rest of the network. In our context, a ``community'' is a group of jobs that share many of the same skills---they form a natural occupational domain.

We applied the \textbf{Louvain algorithm}, which iteratively groups nodes to maximize \textit{modularity}---a measure of how well-separated communities are from each other. The algorithm detected \textbf{162 distinct communities}, ranging from large clusters like Sales/Marketing (2,156 jobs) to specialized niches. Table \ref{tab:communities} presents the eight largest clusters.

\begin{table}[H]
\centering
\caption{Louvain Community Detection Results: Top Occupation Clusters}
\label{tab:communities}
\begin{threeparttable}
\begin{tabular}{@{}c l r c c l@{}}
\toprule
\textbf{ID} & \textbf{Cluster Profile} & \textbf{$|V|$} & \textbf{$\bar{\rho}$} & \textbf{$Q_{\text{int}}$} & \textbf{Sample Occupations} \\
\midrule
7160 & Sales/Marketing & 1,933 & 35.9\% & 0.912 & Sales Exec, Key Account Mgr \\
6167 & Operations/Admin & 1,782 & 37.2\% & 0.887 & QA Engineer, Purchasing Spec \\
15768 & Software/IT & 1,604 & \cellcolor{lowrisk!20}32.7\% & 0.934 & React Dev, Backend Eng \\
2420 & Creative/Design & 1,085 & \cellcolor{medrisk!20}46.9\% & 0.856 & Graphic Designer, PR Spec \\
1819 & HR/Recruitment & 881 & 35.4\% & 0.891 & Talent Acquisition, HR Sup \\
5244 & Finance/Accounting & 755 & \cellcolor{highrisk!20}51.3\% & 0.903 & Financial Analyst, Controller \\
7461 & Admin/Clerical & 429 & \cellcolor{highrisk!20}51.3\% & 0.867 & Executive Secretary, Claims \\
10012 & Hospitality/F\&B & 282 & \cellcolor{lowrisk!20}31.9\% & 0.945 & Sous Chef, Guest Experience \\
\bottomrule
\end{tabular}
\begin{tablenotes}[flushleft]
\footnotesize
\item $|V|$ = community size (jobs + skills); $\bar{\rho}$ = mean automation risk; $Q_{\text{int}}$ = internal modularity (cohesion).
\item Color coding: \colorbox{lowrisk!20}{Low risk ($\le$40\%)}, \colorbox{medrisk!20}{Medium (40--50\%)}, \colorbox{highrisk!20}{High ($>$50\%)}.
\end{tablenotes}
\end{threeparttable}
\end{table}

\textbf{Reading the Table:} The internal modularity $Q_{\text{int}}$ measures how ``tight-knit'' each community is---higher values mean jobs within that cluster share more skills with each other than with outsiders. Software/IT ($Q_{\text{int}} = 0.934$) and Hospitality ($Q_{\text{int}} = 0.945$) are particularly cohesive, meaning their skills are highly specialized and not easily transferable. Conversely, the lower $Q_{\text{int}}$ of Creative/Design (0.856) suggests more skill overlap with adjacent fields.

\textbf{Strategic Implication:} The Finance/Accounting and Admin/Clerical clusters face a double challenge: high automation risk ($\bar{\rho} > 50\%$) \textit{and} high internal cohesion ($Q_{\text{int}} > 0.86$). Workers in these clusters possess skills that are both automatable and not easily transferable---they are ``trapped'' in high-risk domains. In contrast, Software/IT and Hospitality offer ``safe harbors'' with low automation risk, though their high cohesion means entering these fields requires acquiring specialized skills.

\subsection{Bridge Skills Analysis via Betweenness Centrality}
Given that occupational communities are relatively isolated (high modularity), how can workers transition \textit{between} them? The answer lies in \textbf{bridge skills}---activities that appear across multiple communities, creating pathways for cross-sector mobility.

We identify bridge skills using \textit{betweenness centrality}, a network metric that measures how often a node lies on the shortest path between other nodes. Intuitively, a skill with high betweenness centrality acts as a ``gateway''---if you want to move from Finance to IT, you likely need to pass through a bridge skill that both domains require. Formally:
\begin{equation}
C_B(v) = \sum_{s \neq v \neq t} \frac{\sigma_{st}(v)}{\sigma_{st}}
\end{equation}
where $\sigma_{st}$ is the total number of shortest paths from node $s$ to $t$, and $\sigma_{st}(v)$ is the number passing through $v$.

To make this actionable, we developed the \textbf{Connection Pairs} metric $C_p(s)$, which counts how many job-to-job transitions a skill enables across different communities:
\begin{equation}
C_p(s) = \sum_{c_i \neq c_j} |Links(c_i, s)| \cdot |Links(c_j, s)|
\end{equation}
A skill linking 10 jobs in Finance and 20 jobs in IT creates $10 \times 20 = 200$ potential transition pathways. Table \ref{tab:centrality} ranks skills by this bridging power.

\begin{table}[H]
\centering
\caption{Bridge Skills: Betweenness Centrality and Cross-Community Connectivity}
\label{tab:centrality}
\begin{threeparttable}
\small
\begin{tabular}{@{}c l r r c c c@{}}
\toprule
\textbf{\#} & \textbf{Skill Activity} & \textbf{$C_p$} & \textbf{$k$} & \textbf{$D_{\text{ISCO}}$} & \textbf{$\bar{\rho}$} & \textbf{Risk} \\
\midrule
1 & Project Planning \& Scheduling & 332,352 & 1,154 & 15 & 33.0\% & \cellcolor{lowrisk!20}Low \\
2 & Contact Center Customer Service & 279,312 & 1,058 & 27 & 42.3\% & \cellcolor{medrisk!20}Med \\
3 & Sales Team Leadership & 217,622 & 934 & 10 & 28.7\% & \cellcolor{lowrisk!20}Low \\
4 & Business Dev \& Sales Mgmt & 168,510 & 822 & 12 & 30.7\% & \cellcolor{lowrisk!20}Low \\
5 & Quality Engineering Mgmt & 149,382 & 774 & 27 & 34.6\% & \cellcolor{lowrisk!20}Low \\
6 & Regulatory Compliance & 125,670 & 710 & 18 & 33.0\% & \cellcolor{lowrisk!20}Low \\
7 & Client \& Stakeholder Mgmt & 116,622 & 684 & 14 & 29.4\% & \cellcolor{lowrisk!20}Low \\
8 & Financial Market Analysis & 111,890 & 670 & 10 & 38.7\% & \cellcolor{lowrisk!20}Low \\
9 & Warehouse \& Inventory Mgmt & 107,256 & 656 & 22 & 40.8\% & \cellcolor{medrisk!20}Med \\
10 & Team Leadership \& Dev & 95,172 & 618 & 12 & 24.6\% & \cellcolor{lowrisk!20}Low \\
\bottomrule
\end{tabular}
\begin{tablenotes}[flushleft]
\footnotesize
\item $C_p$ = Connection Pairs (transition pathways enabled); $k$ = degree (total jobs requiring this skill); $D_{\text{ISCO}}$ = ISCO-2 diversity (how many occupation categories); $\bar{\rho}$ = mean automation risk of jobs with this skill.
\end{tablenotes}
\end{threeparttable}
\end{table}

\textbf{Reading the Table:} ``Project Planning \& Scheduling'' ranks first with $C_p = 332{,}352$, meaning it enables over 330,000 potential cross-community job transitions. It connects to 1,154 jobs ($k$) across 15 different ISCO-2 occupation categories ($D_{\text{ISCO}}$). Critically, jobs requiring this skill have low average automation risk ($\bar{\rho} = 33.0\%$)---making it both a mobility enabler and a pathway to safety.

Two skills stand out for their \textit{breadth}: ``Contact Center Customer Service'' and ``Quality Engineering Management'' each span \textbf{27 ISCO-2 categories}---the maximum observed. These are ``universal passports'' that provide entry points to nearly every sector of the economy.

\vspace{0.5em}
\noindent\textbf{From Bridging Power to Training Priority.} Not all bridge skills are equally valuable for workforce development. A skill might enable many transitions but lead to high-risk destinations. To identify \textit{optimal} retraining investments, we compute a PageRank-style importance score that balances \textit{frequency} (how many jobs require it) with \textit{diversity} (how many sectors it spans):
\begin{equation}
I_{PR} = n_{\text{jobs}} \times D_{\text{ISCO}}
\end{equation}

Table \ref{tab:pagerank} ranks skills by this importance score, with priority tiers based on both $I_{PR}$ and destination risk $\bar{\rho}$.

\begin{table}[H]
\centering
\caption{Skill Importance via PageRank-Style Algorithm}
\label{tab:pagerank}
\begin{threeparttable}
\small
\begin{tabular}{@{}c l r c r c l@{}}
\toprule
\textbf{\#} & \textbf{Skill} & \textbf{$n_{\text{jobs}}$} & \textbf{$D_{\text{ISCO}}$} & \textbf{$I_{PR}$} & \textbf{$\bar{\rho}$} & \textbf{Priority} \\
\midrule
1 & Contact Center Customer Service & 529 & 27 & 14,283 & 42.3\% & \textbf{Universal} \\
2 & Quality Engineering Mgmt & 387 & 27 & 10,449 & 34.6\% & \textbf{Universal} \\
3 & Project Planning \& Scheduling & 577 & 15 & 8,655 & 33.0\% & Tier~1 \\
4 & Warehouse \& Inventory Mgmt & 328 & 22 & 7,216 & 40.8\% & Tier~2 \\
5 & Regulatory Compliance & 355 & 18 & 6,390 & 33.0\% & Tier~1 \\
6 & Business Reporting & 293 & 21 & 6,153 & 38.9\% & Tier~2 \\
7 & Reference Data Analysis & 321 & 17 & 5,457 & 37.8\% & Tier~2 \\
8 & Business Dev \& Sales Mgmt & 411 & 12 & 4,932 & 30.7\% & Tier~1 \\
9 & Client \& Stakeholder Mgmt & 342 & 14 & 4,788 & 29.4\% & Tier~1 \\
10 & Sales Team Leadership & 467 & 10 & 4,670 & 28.7\% & Tier~1 \\
\bottomrule
\end{tabular}
\begin{tablenotes}[flushleft]
\footnotesize
\item $I_{PR}$ = Importance Score $= n_{\text{jobs}} \times D_{\text{ISCO}}$. Priority tiers: \textbf{Universal} = valuable for all workers; \textbf{Tier~1} = low risk ($\bar{\rho} < 35\%$); \textbf{Tier~2} = moderate risk ($35\% \le \bar{\rho} < 45\%$).
\end{tablenotes}
\end{threeparttable}
\end{table}

\textbf{Policy Implication:} The two ``Universal'' skills---Contact Center Customer Service and Quality Engineering Management---should be priorities for national workforce development programs. They provide the widest range of career options (27 sectors each) while maintaining reasonable automation risk. For workers in high-risk clusters like Finance/Accounting, acquiring these skills offers the most efficient escape route to diverse, safer employment.

\section{Transition Pathways via Graph Traversal}

Given that we have mapped jobs, skills, and automation risk onto a network, a practical question emerges: \textit{which career moves actually reduce a worker's automation exposure?} We answer this by treating job transitions as \textbf{graph traversal}---finding paths from high-risk ``source'' jobs to low-risk ``destination'' jobs, where the path is paved by shared skills.

Our pathfinding algorithm identified \textbf{4,534 realistic transitions} satisfying our dual threshold criteria: (1) the destination is safer than the origin ($\Delta \rho < 0$), (2) the jobs share at least 3 activities ($|\mathcal{A}_{shared}| \ge 3$), and (3) at least 50\% of the source job's skills transfer to the destination. This rigorous threshold ensures that identified transitions leverage existing competencies rather than requiring near-complete retraining. Table \ref{tab:transition_stats} quantifies the resulting transition network.

\begin{table}[H]
\centering
\caption{Transition Network Statistics from Graph Pathfinding Algorithm (Dual Threshold)}
\label{tab:transition_stats}
\begin{threeparttable}
\begin{tabular}{@{}l l r@{}}
\toprule
\textbf{Category} & \textbf{Metric} & \textbf{Value} \\
\midrule
\multirow{4}{*}[0pt]{\textbf{Path Statistics}}
& Total Realistic Transitions ($|T|$) & 4,534 \\
& Mean Shared Skills ($\bar{k}_{\text{shared}}$) & 3.8 \\
& Max Shared Skills & 8 \\
& Mean Skill Transfer Rate ($\bar{\phi}$) & 53.5\% \\
\addlinespace[3pt]
\multirow{3}{*}[0pt]{\textbf{Source Nodes}}
& Unique High-Risk Sources & 509 \\
& Mean Out-Degree (transitions available) & 8.9 \\
& Sources with ${>}10$ Options & 89 (17.5\%) \\
\addlinespace[3pt]
\multirow{3}{*}[0pt]{\textbf{Destination Nodes}}
& Unique Low-Risk Destinations & 1,684 \\
& Mean In-Degree (accessibility) & 2.7 \\
& Hub Destinations ($k_{\text{in}} > 100$) & 0 \\
\addlinespace[3pt]
\multirow{2}{*}[0pt]{\textbf{Risk Reduction}}
& Mean Risk Reduction ($\bar{\Delta\rho}$) & $-$48.1~pp \\
& Max Risk Reduction & $-$89.0~pp \\
\addlinespace[3pt]
\multirow{2}{*}[0pt]{\textbf{Coverage Gap}}
& High-Risk Jobs with Paths & 509 / 2,089 (24.4\%) \\
& High-Risk Jobs Needing Reskilling & 1,580 / 2,089 (75.6\%) \\
\bottomrule
\end{tabular}
\begin{tablenotes}[flushleft]
\footnotesize
\item pp = percentage points. Skill transfer rate = shared skills / source job total skills.
\end{tablenotes}
\end{threeparttable}
\end{table}

\textbf{Reading the Table:} The transition network connects 509 high-risk ``source'' jobs to 1,684 low-risk ``destinations'' via 4,534 pathways. The \textbf{coverage gap} is the critical finding: only 24.4\% of high-risk workers have realistic organic transition pathways---the remaining 75.6\% (1,580 workers) require substantial reskilling interventions.

For those with viable paths, the quality is high: mean skill transfer of 53.5\% means workers leverage the \textit{majority} of their existing competencies. The mean risk reduction of 48.1 percentage points indicates that successful transitions approximately halve automation exposure. The mean out-degree of 8.9 options per worker and absence of hub destinations underscore that while pathways are more abundant than previously estimated, they remain specific---there is no ``one-size-fits-all'' safe harbor.

\subsection{Safe Harbor Analysis via In-Degree Centrality}
If many workers need to transition, \textit{where should they go?} Some occupations act as natural ``sinks'' for displaced talent---jobs that are both low-risk \textit{and} reachable from many high-risk starting points. We identify these \textbf{Safe Harbors} using in-degree centrality ($k_{in}$): the count of distinct high-risk jobs from which a destination is accessible. Table \ref{tab:safe_harbor} ranks the top safe harbors.

\begin{table}[H]
\centering
\caption{Safe Harbor Destinations: In-Degree Centrality and Accessibility Metrics}
\label{tab:safe_harbor}
\begin{threeparttable}
\small
\begin{tabular}{@{}c l c r c c r@{}}
\toprule
\textbf{\#} & \textbf{Destination Job} & \textbf{$\rho$} & \textbf{$k_{\text{in}}$} & \textbf{$\bar{J}$} & \textbf{$|\mathcal{A}|$} & \textbf{Bridge} \\
\midrule
1 & Administrative Manager & \cellcolor{lowrisk!20}29.0\% & 498 & 0.51 & 40 & 2,794 \\
2 & Branch Manager & \cellcolor{lowrisk!20}30.0\% & 452 & 0.48 & 21 & 2,223 \\
3 & Supply \& Demand Planning Lead & \cellcolor{lowrisk!20}31.3\% & 442 & 0.47 & 20 & 2,466 \\
4 & SPA Manager & \cellcolor{verylowrisk!20}22.6\% & 438 & 0.45 & 35 & 1,892 \\
5 & Brand Supervisor & \cellcolor{lowrisk!20}34.7\% & 437 & 0.52 & 40 & 2,794 \\
6 & Facilities Manager & \cellcolor{lowrisk!20}31.0\% & 432 & 0.49 & 40 & 2,302 \\
7 & Asst Chief Steward & \cellcolor{lowrisk!20}37.5\% & 429 & 0.44 & 28 & 1,756 \\
8 & Branch Manager (Retail) & \cellcolor{lowrisk!20}32.4\% & 423 & 0.46 & 21 & 2,223 \\
9 & HR Service Delivery Manager & \cellcolor{verylowrisk!20}\textbf{4.3\%} & 418 & 0.43 & 32 & 1,834 \\
10 & Finance Manager (Ops) & \cellcolor{lowrisk!20}32.1\% & 412 & 0.48 & 38 & 2,180 \\
\bottomrule
\end{tabular}
\begin{tablenotes}[flushleft]
\footnotesize
\item $k_{\text{in}}$ = in-degree (accessibility); $\bar{J}$ = mean Jaccard similarity; $|\mathcal{A}|$ = activity count; Bridge = connected jobs.
\item Color: \colorbox{verylowrisk!20}{Very low ($<$25\%)}, \colorbox{lowrisk!20}{Low (25--40\%)}.
\end{tablenotes}
\end{threeparttable}
\end{table}

\textbf{Reading the Table:} The column $k_{\text{in}}$ shows how many high-risk jobs can reach each destination. ``Administrative Manager'' ranks first ($k_{in} = 498$)---nearly 500 different at-risk occupations share enough skills to make this transition viable. The Jaccard similarity ($\bar{J} = 0.51$) confirms that these aren't distant leaps; on average, workers already possess half the required skills. The ``Bridge'' column shows total job connections, indicating how central each role is in the broader labor market.

\textbf{Strategic Insight:} ``HR Service Delivery Manager'' presents an exceptional opportunity: ultra-low automation risk ($\rho = 4.3\%$) combined with high accessibility ($k_{in} = 418$). This role should be a priority target for workforce transition programs---it offers maximum safety with minimal retraining barriers.

\subsection{Skill Gap Analysis: The Process Skills Multiplier}
Knowing \textit{where} to transition is only half the answer. The next question is: \textit{what skills must workers acquire to get there?} We analyze ``gap skills''---competencies required by the destination job but absent from the source. Table \ref{tab:skill_gaps} ranks skills by how frequently they appear as gaps across all 4,534 realistic transitions.

\begin{table}[H]
\centering
\caption{Skill Gaps for Realistic Transitions: Frequency Analysis (Dual Threshold)}
\label{tab:skill_gaps}
\begin{threeparttable}
\begin{tabular}{@{}c l r r r@{}}
\toprule
\textbf{\#} & \textbf{Skill to Acquire} & \textbf{$f_{\text{gap}}$} & \textbf{\% of $|T|$} & \textbf{Cum.~\%} \\
\midrule
\rowcolor{medrisk!15} 1 & \textbf{Process Improvement} & \textbf{708} & \textbf{15.6\%} & 15.6\% \\
2 & Custom Report Generation & 642 & 14.2\% & 29.8\% \\
3 & Operations Team Coordination & 629 & 13.9\% & 43.7\% \\
4 & Budget Management & 610 & 13.5\% & 57.2\% \\
5 & Requirements Analysis & 592 & 13.1\% & 70.3\% \\
\midrule
6 & Project Planning & 592 & 13.1\% & --- \\
7 & Employee Training Delivery & 558 & 12.3\% & --- \\
8 & Performance Monitoring & 543 & 12.0\% & --- \\
9 & Risk Assessment & 533 & 11.8\% & --- \\
10 & Quality Assurance & 473 & 10.4\% & --- \\
\bottomrule
\end{tabular}
\begin{tablenotes}[flushleft]
\footnotesize
\item $f_{\text{gap}}$ = frequency as gap skill across all realistic transitions ($|T| = 4{,}534$).
\item Skills overlap across pathways, so cumulative percentages indicate coverage rather than sum to 100\%.
\end{tablenotes}
\end{threeparttable}
\end{table}

\textbf{Reading the Table:} The column $f_{\text{gap}}$ counts how many transitions require each skill. ``Process Improvement'' tops the list at 708---meaning this single skill appears as a gap in \textbf{15.6\%} of all realistic transitions. The concentration is substantial: the top 5 skills alone address over 70\% of all pathways.

\textbf{The Process Skills Multiplier:} This finding has profound policy implications. Consider a typical clerical worker facing high automation risk ($\rho \ge 60\%$). Among the 24.4\% of high-risk workers who have realistic transition paths, the most common barriers are process-oriented skills---Process Improvement, Custom Report Generation, and Operations Team Coordination. A single, well-designed intervention in process optimization can unlock the largest number of safe career destinations. This is the ``process skills multiplier'': one skill investment yields disproportionate mobility returns.

\textbf{The Reskilling Imperative:} For the 75.6\% of high-risk workers \textit{without} realistic organic pathways, gap skill training alone is insufficient. These workers require comprehensive reskilling programs that build entirely new competency foundations. The skill gap table above identifies priority training areas even for this cohort: Process Improvement, Budget Management, and Project Planning represent capabilities that expand access to safer occupations across multiple ISCO categories.

\subsection{Exemplar Transition Pathways: From Theory to Action}
The preceding analysis identifies \textit{aggregate} patterns---but what does a safe transition actually look like for an individual worker? To make our methodology concrete, we present five detailed case studies. Each pathway was selected for: (1) substantial risk reduction, (2) relatable source occupations common in Egypt, and (3) clear skill bridges that illustrate how shared competencies enable mobility.

\begin{table}[H]
\centering
\caption{Exemplar Career Transition Pathways: Detailed Skill Analysis}
\label{tab:exemplar_transitions}
\begin{threeparttable}
\footnotesize
\begin{tabular}{@{}p{2.8cm} p{2.8cm} c c c c@{}}
\toprule
\textbf{Source Job} & \textbf{Target Job} & \textbf{$\rho_s$} & \textbf{$\rho_t$} & \textbf{$\Delta\rho$} & \textbf{$J$} \\
\midrule
\rowcolor{highrisk!10}
Data Entry Clerk & Administrative Manager & 78.5\% & 29.0\% & $-$49.5 & 12\% \\
\rowcolor{highrisk!10}
Sales Executive & Account Manager & 80.0\% & 20.0\% & $-$60.0 & 31\% \\
\rowcolor{medrisk!10}
Procurement Buyer & Sales Manager & 73.3\% & 16.0\% & $-$57.3 & 18\% \\
\rowcolor{medrisk!10}
Medical Representative & Senior Medical Rep & 68.0\% & 16.7\% & $-$51.3 & 42\% \\
\rowcolor{medrisk!10}
Marketing Strategist & Brand Manager & 71.0\% & 39.0\% & $-$32.0 & 35\% \\
\bottomrule
\end{tabular}
\begin{tablenotes}[flushleft]
\scriptsize
\item $\rho_s$ = source risk; $\rho_t$ = target risk; $\Delta\rho$ = risk reduction (pp); $J$ = Jaccard similarity.
\end{tablenotes}
\end{threeparttable}
\end{table}

\textbf{Reading the Table:} Each row represents a real transition pathway from our graph. The risk reduction ($\Delta\rho$) shows the ``safety gain''---for instance, a Sales Executive moving to Account Manager reduces automation exposure by 60 percentage points. The Jaccard similarity ($J$) indicates skill overlap: higher values mean easier transitions. Note that even the ``hardest'' transition here (Data Entry $\rightarrow$ Admin Manager, $J = 12\%$) is still viable because the jobs share sufficient core competencies.

Table \ref{tab:transition_details} unpacks what these transitions require in practice---the shared skills that make the pathway possible, and the gap skills workers must acquire.

\begin{table}[H]
\centering
\caption{Detailed Skill Decomposition for Exemplar Transitions}
\label{tab:transition_details}
\begin{threeparttable}
\scriptsize
\begin{tabular}{@{}p{1.8cm} p{4.8cm} p{4.8cm} c@{}}
\toprule
\textbf{Transition} & \textbf{Shared Skills (Bridge)} & \textbf{Gap Skills (To Acquire)} & \textbf{$n_{\text{gap}}$} \\
\midrule
\rowcolor{lightgray}
Data Entry $\rightarrow$ Admin Mgr &
Document Management, Office Administration, Report Generation, Data Verification, Internal Communication &
\textbf{Team Leadership}, Budget Management, Strategic Planning, Stakeholder Management, Performance Evaluation &
35 \\
\addlinespace[2pt]
Sales Exec $\rightarrow$ Account Mgr &
Client Relationship Mgmt, Sales Presentations, CRM Operations, Lead Generation, Market Analysis &
\textbf{Account Strategy}, Contract Negotiation, Revenue Forecasting, Key Account Planning &
12 \\
\addlinespace[2pt]
\rowcolor{lightgray}
Procurement $\rightarrow$ Sales Mgr &
Vendor Negotiation, Contract Management, Supply Chain Knowledge, Cost Analysis &
\textbf{Sales Team Leadership}, Pipeline Management, Territory Planning, Quota Setting &
18 \\
\addlinespace[2pt]
Med Rep $\rightarrow$ Sr Med Rep &
Product Knowledge, Healthcare Provider Relations, Compliance, Territory Management, Sales Reporting &
\textbf{Team Mentoring}, Training Delivery, Regional Strategy, KOL Management &
8 \\
\addlinespace[2pt]
\rowcolor{lightgray}
Marketing $\rightarrow$ Brand Mgr &
Market Research, Campaign Planning, Brand Positioning, Consumer Insights, Digital Marketing &
\textbf{P\&L Ownership}, Product Lifecycle Mgmt, Cross-functional Leadership, Agency Management &
14 \\
\bottomrule
\end{tabular}
\begin{tablenotes}[flushleft]
\scriptsize
\item $n_{\text{gap}}$ = total gap skills needed. \textbf{Bold} indicates highest-priority gap skill for each transition.
\item All transitions satisfy dual threshold: $|\mathcal{A}_{\text{shared}}| \ge 3$ AND skill transfer $\ge 50\%$.
\end{tablenotes}
\end{threeparttable}
\end{table}

\begin{figure}[H]
\centering
\begin{tikzpicture}[
    scale=0.85,
    transform shape,
    sourcejob/.style={circle, draw=red!80, fill=red!40, minimum size=55pt, inner sep=1pt, font=\scriptsize\bfseries, text width=1.5cm, align=center, line width=1.5pt},
    targetjob/.style={circle, draw=green!70!black, fill=green!35, minimum size=55pt, inner sep=1pt, font=\scriptsize\bfseries, text width=1.5cm, align=center, line width=1.5pt},
    sharedact/.style={circle, draw=teal!80, fill=teal!30, minimum size=14pt, inner sep=0pt, line width=1.2pt},
    unusedact/.style={circle, draw=gray!60, fill=gray!25, minimum size=14pt, inner sep=0pt, line width=1pt},
    gapact/.style={circle, draw=orange!80, fill=orange!40, minimum size=14pt, inner sep=0pt, line width=1.2pt},
    sharedtool/.style={circle, draw=teal!90, fill=teal!55, minimum size=7pt, inner sep=0pt, line width=0.8pt},
    unusedtool/.style={circle, draw=gray!80, fill=gray!50, minimum size=7pt, inner sep=0pt, line width=0.8pt},
    gaptool/.style={circle, draw=orange!90, fill=orange!60, minimum size=7pt, inner sep=0pt, line width=0.8pt},
    performsshared/.style={draw=teal!60, line width=1.3pt, opacity=0.85},
    performsunused/.style={draw=gray!50, line width=1pt, opacity=0.6},
    performsgap/.style={draw=orange!60, line width=1.3pt, opacity=0.85},
    usesedge/.style={draw=gray!40, line width=0.8pt, opacity=0.7},
    transarrow/.style={-{Stealth[length=12pt]}, draw=purple!70, line width=3pt},
]

\node[sourcejob] (src) at (0,0) {Data Entry Clerk};
\node[font=\scriptsize, text=red!70!black] at (0,-1.6) {$\rho = 78.5\%$};

\node[targetjob] (tgt) at (11,0) {Admin Manager};
\node[font=\scriptsize, text=green!50!black] at (11,-1.6) {$\rho = 29.0\%$};

\draw[transarrow] (1.8,0) -- (9.2,0);
\node[font=\footnotesize\bfseries, text=purple!70!black, fill=white, inner sep=4pt] at (5.5,0) {$\Delta\rho = -49.5$ pp};

\node[sharedact, label={[font=\tiny, text=teal!80!black]above:Document Mgmt}] (sh1) at (3.5,2.8) {};
\node[sharedact, label={[font=\tiny, text=teal!80!black]above:Office Admin}] (sh2) at (5.5,3.2) {};
\node[sharedact, label={[font=\tiny, text=teal!80!black]above:Report Gen}] (sh3) at (7.5,2.8) {};
\node[sharedact, label={[font=\tiny, text=teal!80!black]below:Data Verify}] (sh4) at (4.2,1.6) {};
\node[sharedact, label={[font=\tiny, text=teal!80!black]below:Internal Comm}] (sh5) at (6.8,1.6) {};

\draw[performsshared] (src) to[bend left=25] (sh1);
\draw[performsshared] (src) to[bend left=35] (sh2);
\draw[performsshared] (src) to[bend left=20] (sh3);
\draw[performsshared] (src) to[bend left=12] (sh4);
\draw[performsshared] (src) to[bend left=8] (sh5);

\draw[performsshared] (sh1) to[bend left=25] (tgt);
\draw[performsshared] (sh2) to[bend left=35] (tgt);
\draw[performsshared] (sh3) to[bend left=20] (tgt);
\draw[performsshared] (sh4) to[bend left=8] (tgt);
\draw[performsshared] (sh5) to[bend left=12] (tgt);

\node[unusedact, label={[font=\tiny, text=gray!70]below:Data Entry}] (un1) at (-0.5,-2.8) {};
\node[unusedact, label={[font=\tiny, text=gray!70]below:Form Processing}] (un2) at (1.5,-2.8) {};

\draw[performsunused] (src) to[bend right=25] (un1);
\draw[performsunused] (src) to[bend right=20] (un2);

\node[gapact, label={[font=\tiny\bfseries, text=orange!80!black]below:Team Leadership}] (gp1) at (9.5,-2.5) {};
\node[gapact, label={[font=\tiny, text=orange!80!black]below:Budget Mgmt}] (gp2) at (11.5,-2.5) {};
\node[gapact, label={[font=\tiny, text=orange!80!black]below:Strategic Plan}] (gp3) at (13,-1.8) {};
\node[gapact, label={[font=\tiny, text=orange!80!black]above:Stakeholder Mgmt}] (gp4) at (13,0.8) {};
\node[gapact, label={[font=\tiny, text=orange!80!black]above:Perf Evaluation}] (gp5) at (13,-0.5) {};

\draw[performsgap] (tgt) to[bend right=25] (gp1);
\draw[performsgap] (tgt) to[bend right=30] (gp2);
\draw[performsgap] (tgt) to[bend right=15] (gp3);
\draw[performsgap] (tgt) to[bend left=15] (gp4);
\draw[performsgap] (tgt) to[bend left=5] (gp5);

\node[sharedtool, label={[font=\tiny, text=teal!60!black]right:MS Word}] (t1) at (4.2,3.8) {};
\node[sharedtool, label={[font=\tiny, text=teal!60!black]left:Excel}] (t2) at (6.8,4.0) {};
\node[sharedtool, label={[font=\tiny, text=teal!60!black]right:Email}] (t3) at (5.5,1.0) {};
\draw[usesedge] (sh1) -- (t1);
\draw[usesedge] (sh2) -- (t2);
\draw[usesedge] (sh3) -- (t2);
\draw[usesedge] (sh5) -- (t3);

\node[unusedtool, label={[font=\tiny, text=gray!60]left:Data Entry SW}] (t4) at (-1.5,-3.5) {};
\draw[usesedge] (un1) -- (t4);

\node[gaptool, label={[font=\tiny, text=orange!60!black]right:ERP System}] (t5) at (14,0.2) {};
\node[gaptool, label={[font=\tiny, text=orange!60!black]right:Budget SW}] (t6) at (12.8,-3.2) {};
\draw[usesedge] (gp2) -- (t6);
\draw[usesedge] (gp3) -- (t5);
\draw[usesedge] (gp4) -- (t5);

\draw[gray!30, line width=0.6pt, rounded corners=6pt, fill=gray!5] (-2,-5.2) rectangle (14.5,-4.0);
\node[font=\tiny\bfseries] at (-0.5,-4.3) {Activities (medium):};
\node[sharedact, minimum size=8pt] at (1.6,-4.3) {};
\node[font=\tiny] at (2.8,-4.3) {Shared};
\node[unusedact, minimum size=8pt] at (4.0,-4.3) {};
\node[font=\tiny] at (5.1,-4.3) {Unused};
\node[gapact, minimum size=8pt] at (6.2,-4.3) {};
\node[font=\tiny] at (7.0,-4.3) {Gap};
\node[font=\tiny\bfseries] at (-0.5,-4.85) {Tools (small):};
\node[sharedtool, minimum size=5pt] at (1.6,-4.85) {};
\node[font=\tiny] at (2.8,-4.85) {Shared};
\node[unusedtool, minimum size=5pt] at (4.0,-4.85) {};
\node[font=\tiny] at (5.1,-4.85) {Unused};
\node[gaptool, minimum size=5pt] at (6.2,-4.85) {};
\node[font=\tiny] at (7.0,-4.85) {New};
\draw[performsshared, line width=1pt] (8.2,-4.3) -- (8.9,-4.3);
\node[font=\tiny] at (10.0,-4.3) {PERFORMS};
\draw[usesedge] (11.2,-4.3) -- (11.9,-4.3);
\node[font=\tiny] at (12.7,-4.3) {USES};

\end{tikzpicture}
\caption{\textbf{Transition Pathway 1: Data Entry Clerk $\rightarrow$ Administrative Manager} (Jaccard = 12\%). All nodes are \textbf{circles} distinguished by size: \textbf{large} = jobs, \textbf{medium} = activities, \textbf{small} = tools (darker shades). Color coding: \textbf{Teal} = shared/transferable (the bridge), \textbf{Gray} = unused source elements, \textbf{Orange} = gap elements to acquire through training.}
\label{fig:transition1}
\end{figure}
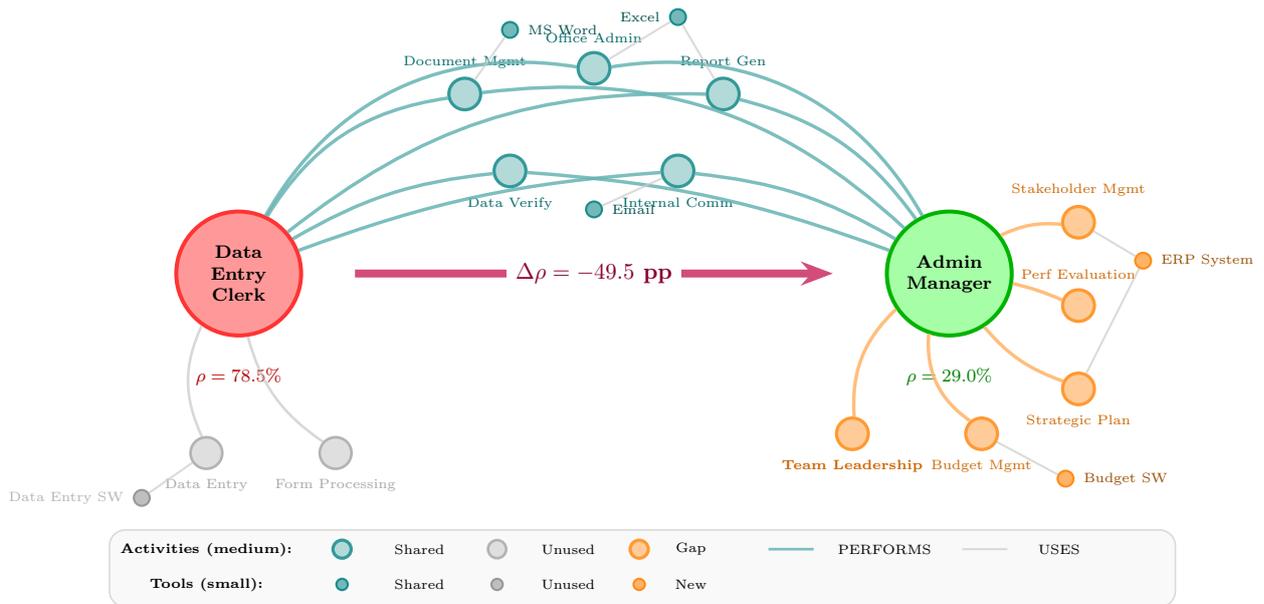

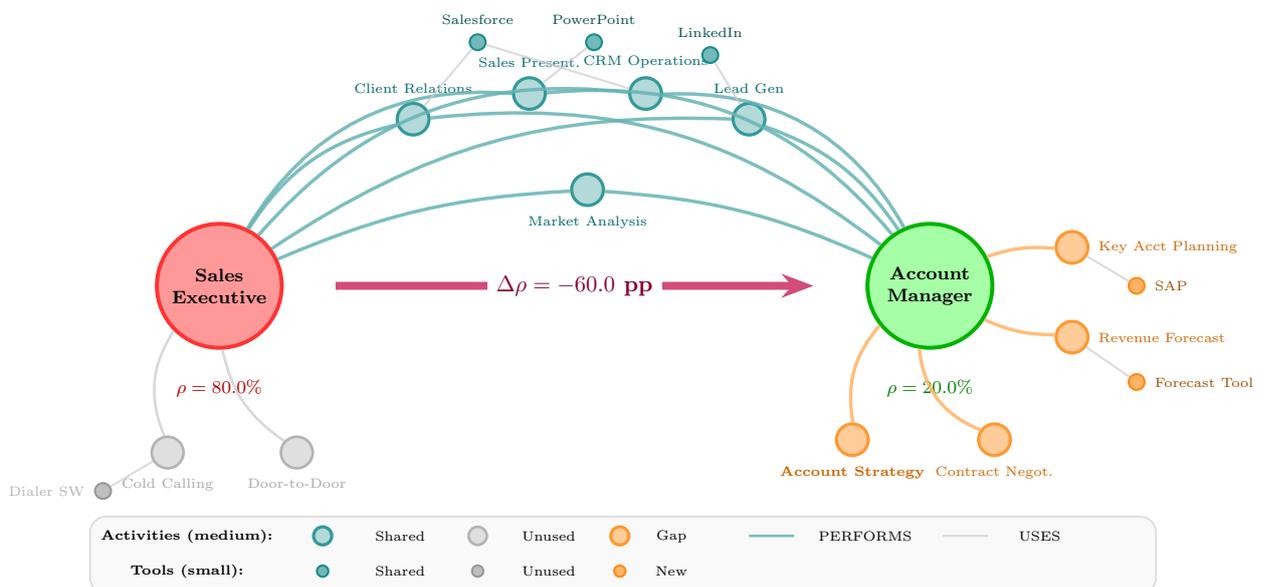
\begin{figure}[H]
\centering
\begin{tikzpicture}[
    scale=0.85,
    transform shape,
    sourcejob/.style={circle, draw=red!80, fill=red!40, minimum size=55pt, inner sep=1pt, font=\scriptsize\bfseries, text width=1.5cm, align=center, line width=1.5pt},
    targetjob/.style={circle, draw=green!70!black, fill=green!35, minimum size=55pt, inner sep=1pt, font=\scriptsize\bfseries, text width=1.5cm, align=center, line width=1.5pt},
    sharedact/.style={circle, draw=teal!80, fill=teal!30, minimum size=14pt, inner sep=0pt, line width=1.2pt},
    unusedact/.style={circle, draw=gray!60, fill=gray!25, minimum size=14pt, inner sep=0pt, line width=1pt},
    gapact/.style={circle, draw=orange!80, fill=orange!40, minimum size=14pt, inner sep=0pt, line width=1.2pt},
    sharedtool/.style={circle, draw=teal!90, fill=teal!55, minimum size=7pt, inner sep=0pt, line width=0.8pt},
    unusedtool/.style={circle, draw=gray!80, fill=gray!50, minimum size=7pt, inner sep=0pt, line width=0.8pt},
    gaptool/.style={circle, draw=orange!90, fill=orange!60, minimum size=7pt, inner sep=0pt, line width=0.8pt},
    performsshared/.style={draw=teal!60, line width=1.3pt, opacity=0.85},
    performsunused/.style={draw=gray!50, line width=1pt, opacity=0.6},
    performsgap/.style={draw=orange!60, line width=1.3pt, opacity=0.85},
    usesedge/.style={draw=gray!40, line width=0.8pt, opacity=0.7},
    transarrow/.style={-{Stealth[length=12pt]}, draw=purple!70, line width=3pt},
]

\node[sourcejob] (src) at (0,0) {Sales Executive};
\node[font=\scriptsize, text=red!70!black] at (0,-1.6) {$\rho = 80.0\%$};

\node[targetjob] (tgt) at (11,0) {Account Manager};
\node[font=\scriptsize, text=green!50!black] at (11,-1.6) {$\rho = 20.0\%$};

\draw[transarrow] (1.8,0) -- (9.2,0);
\node[font=\footnotesize\bfseries, text=purple!70!black, fill=white, inner sep=4pt] at (5.5,0) {$\Delta\rho = -60.0$ pp};

\node[sharedact, label={[font=\tiny, text=teal!80!black]above:Client Relations}] (sh1) at (3.0,2.6) {};
\node[sharedact, label={[font=\tiny, text=teal!80!black]above:Sales Present.}] (sh2) at (4.8,3.0) {};
\node[sharedact, label={[font=\tiny, text=teal!80!black]above:CRM Operations}] (sh3) at (6.6,3.0) {};
\node[sharedact, label={[font=\tiny, text=teal!80!black]above:Lead Gen}] (sh4) at (8.2,2.6) {};
\node[sharedact, label={[font=\tiny, text=teal!80!black]below:Market Analysis}] (sh5) at (5.7,1.5) {};

\draw[performsshared] (src) to[bend left=22] (sh1);
\draw[performsshared] (src) to[bend left=32] (sh2);
\draw[performsshared] (src) to[bend left=28] (sh3);
\draw[performsshared] (src) to[bend left=18] (sh4);
\draw[performsshared] (src) to[bend left=10] (sh5);

\draw[performsshared] (sh1) to[bend left=22] (tgt);
\draw[performsshared] (sh2) to[bend left=28] (tgt);
\draw[performsshared] (sh3) to[bend left=32] (tgt);
\draw[performsshared] (sh4) to[bend left=18] (tgt);
\draw[performsshared] (sh5) to[bend left=10] (tgt);

\node[unusedact, label={[font=\tiny, text=gray!70]below:Cold Calling}] (un1) at (-0.8,-2.6) {};
\node[unusedact, label={[font=\tiny, text=gray!70]below:Door-to-Door}] (un2) at (1.2,-2.6) {};

\draw[performsunused] (src) to[bend right=28] (un1);
\draw[performsunused] (src) to[bend right=22] (un2);

\node[gapact, label={[font=\tiny\bfseries, text=orange!80!black]below:Account Strategy}] (gp1) at (9.8,-2.4) {};
\node[gapact, label={[font=\tiny, text=orange!80!black]below:Contract Negot.}] (gp2) at (12,-2.4) {};
\node[gapact, label={[font=\tiny, text=orange!80!black]right:Revenue Forecast}] (gp3) at (13.2,-0.8) {};
\node[gapact, label={[font=\tiny, text=orange!80!black]right:Key Acct Planning}] (gp4) at (13.2,0.6) {};

\draw[performsgap] (tgt) to[bend right=25] (gp1);
\draw[performsgap] (tgt) to[bend right=32] (gp2);
\draw[performsgap] (tgt) to[bend right=12] (gp3);
\draw[performsgap] (tgt) to[bend left=12] (gp4);

\node[sharedtool, label={[font=\tiny, text=teal!60!black]above:Salesforce}] (t1) at (4.0,3.8) {};
\node[sharedtool, label={[font=\tiny, text=teal!60!black]above:PowerPoint}] (t2) at (5.8,3.8) {};
\node[sharedtool, label={[font=\tiny, text=teal!60!black]above:LinkedIn}] (t3) at (7.6,3.6) {};
\draw[usesedge] (sh1) -- (t1);
\draw[usesedge] (sh2) -- (t2);
\draw[usesedge] (sh3) -- (t1);
\draw[usesedge] (sh4) -- (t3);

\node[unusedtool, label={[font=\tiny, text=gray!60]left:Dialer SW}] (t4) at (-1.8,-3.2) {};
\draw[usesedge] (un1) -- (t4);

\node[gaptool, label={[font=\tiny, text=orange!60!black]right:SAP}] (t5) at (14.2,0) {};
\node[gaptool, label={[font=\tiny, text=orange!60!black]right:Forecast Tool}] (t6) at (14.2,-1.5) {};
\draw[usesedge] (gp3) -- (t6);
\draw[usesedge] (gp4) -- (t5);

\draw[gray!30, line width=0.6pt, rounded corners=6pt, fill=gray!5] (-2,-4.8) rectangle (14.5,-3.6);
\node[font=\tiny\bfseries] at (-0.5,-3.9) {Activities (medium):};
\node[sharedact, minimum size=8pt] at (1.6,-3.9) {};
\node[font=\tiny] at (2.8,-3.9) {Shared};
\node[unusedact, minimum size=8pt] at (4.0,-3.9) {};
\node[font=\tiny] at (5.1,-3.9) {Unused};
\node[gapact, minimum size=8pt] at (6.2,-3.9) {};
\node[font=\tiny] at (7.0,-3.9) {Gap};
\node[font=\tiny\bfseries] at (-0.5,-4.45) {Tools (small):};
\node[sharedtool, minimum size=5pt] at (1.6,-4.45) {};
\node[font=\tiny] at (2.8,-4.45) {Shared};
\node[unusedtool, minimum size=5pt] at (4.0,-4.45) {};
\node[font=\tiny] at (5.1,-4.45) {Unused};
\node[gaptool, minimum size=5pt] at (6.2,-4.45) {};
\node[font=\tiny] at (7.0,-4.45) {New};
\draw[performsshared, line width=1pt] (8.2,-3.9) -- (8.9,-3.9);
\node[font=\tiny] at (10.0,-3.9) {PERFORMS};
\draw[usesedge] (11.2,-3.9) -- (11.9,-3.9);
\node[font=\tiny] at (12.7,-3.9) {USES};

\end{tikzpicture}
\caption{\textbf{Transition Pathway 2: Sales Executive $\rightarrow$ Account Manager} (Jaccard = 31\%). All nodes are \textbf{circles} distinguished by size: \textbf{large} = jobs, \textbf{medium} = activities, \textbf{small} = tools (darker shades). This easier transition has more shared activities relative to gaps (5:4 ratio), reflecting higher skill overlap. Color coding: \textbf{Teal} = shared/transferable, \textbf{Gray} = unused, \textbf{Orange} = gap elements to acquire.}
\label{fig:transition2}
\end{figure}

Figures \ref{fig:transition1} and \ref{fig:transition2} visualize two representative transitions, illustrating how shared skills bridge high-risk and low-risk occupations.

\textbf{Key Insights from Case Studies:}
\begin{enumerate}
    \item \textbf{Process skills are the universal gap.} In the majority of transitions, process-oriented skills (Process Improvement, Operations Coordination, Project Planning) appear as primary gaps. This reinforces the ``process skills multiplier''---investing in operational training yields the highest ROI for workforce mobility.

    \item \textbf{Jaccard similarity predicts training effort.} Higher $J$ means fewer skills to acquire. Medical Rep $\rightarrow$ Senior Med Rep ($J = 42\%$) needs only 8 gap skills; Data Entry $\rightarrow$ Admin Manager ($J = 12\%$) needs 35. Policymakers can use $J$ to estimate program duration and cost.

    \item \textbf{Cross-sector mobility is achievable.} The Procurement $\rightarrow$ Sales Manager path crosses ISCO-2 categories (Business Services to Managers), proving that transferable skills like negotiation and contract management can bridge seemingly distant occupations.

    \item \textbf{Vertical progression inherently reduces risk.} All five transitions yield substantial safety gains (mean: 50 pp reduction), validating our earlier finding that moving ``up'' the occupational hierarchy naturally moves workers away from automatable tasks.
\end{enumerate}

\section{Discussion}

\subsection{Interpretation of Key Findings}
Our analysis reveals several patterns that both confirm and extend existing labor market research. The overall finding that 20.9\% of Egyptian jobs face high automation risk aligns with \citet{arntz2016risk}'s task-based estimates for OECD countries (median 9\%, range 6--12\% for high risk), though our higher figure reflects Egypt's economic structure with greater reliance on routine-intensive occupations.

The \textbf{occupational hierarchy gradient}---where automation risk decreases systematically from clerical (54.6\%) to managerial (30.3\%) roles---supports \citet{autor2015why}'s polarization hypothesis. However, the Egyptian context reveals important nuances: the relatively small proportion of middle-skill manufacturing jobs in our sample (ISCO-7 and ISCO-8 combined represent only 2.1\% of postings) suggests that Egypt's automation challenge is primarily a \textit{white-collar} phenomenon, distinct from the manufacturing-centered disruption observed in advanced economies.

\subsection{The Bridge Skills Mechanism}
Our identification of 25 high-centrality "Bridge Skills" provides empirical support for theoretical arguments about skill complementarity \citep{deming2017growing}. Under our rigorous dual threshold ($\geq$3 shared skills AND $\geq$50\% transfer), "Process Improvement" appears as a gap skill in 708 of 4,534 realistic transitions (15.6\%)---the highest concentration of any single competency. This suggests that process-oriented skills may be systematically undervalued in traditional human capital frameworks that emphasize technical competencies.

Critically, bridge skills exhibit a \textbf{cross-community structure}: they connect occupational clusters that would otherwise be isolated. "Quality Engineering Management," spanning 27 ISCO-2 categories, exemplifies this pattern. From a network resilience perspective, workers possessing bridge skills are less vulnerable to sector-specific shocks because they can traverse the labor market graph through multiple pathways.

\subsection{The Coverage Gap: A Central Finding}
Perhaps the most significant insight from our analysis is the \textbf{coverage gap}: under realistic transition criteria, only 24.4\% of high-risk workers (509 out of 2,089) have organic pathways to safety. The remaining 75.6\% (1,580 workers) face a structural barrier---their current skill profiles do not overlap sufficiently with any safer occupation to enable a low-friction transition.

This finding has been obscured in prior studies using permissive thresholds. With a simple $\tau \geq 3$ criterion, we identified 65,643 theoretical pathways covering 61.3\% of high-risk workers. However, these pathways averaged only 35.7\% skill transfer---meaning workers would need to acquire roughly 65\% new competencies, which is functionally equivalent to career reinvention rather than transition.

The coverage gap is not a methodological limitation but a \textbf{structural reality} of the Egyptian labor market. It underscores the inadequacy of passive "skill matching" approaches and demands active policy intervention to create new pathways where organic ones do not exist.

\subsection{Comparison with International Findings}
Table \ref{tab:international} contextualizes our findings within the broader literature on automation risk.

\begin{table}[H]
\centering
\caption{Comparison of Automation Risk Estimates Across Studies}
\label{tab:international}
\begin{threeparttable}
\begin{tabular}{@{}l l c l@{}}
\toprule
\textbf{Study} & \textbf{Country/Region} & \textbf{High Risk} & \textbf{Method} \\
\midrule
\citet{frey2017future} & United States & 47\% & Occupation-level \\
\citet{arntz2016risk} & OECD Average & 9\% & Task-level \\
\citet{eloundou2023gpts} & United States & 19\%\tnote{a} & LLM exposure \\
\rowcolor{execblue!10} \textbf{This Study} & \textbf{Egypt} & \textbf{20.9\%} & \textbf{Task-level (validated)} \\
\bottomrule
\end{tabular}
\begin{tablenotes}[flushleft]
\footnotesize
\item[a] Percentage of workforce with ${>}50\%$ task exposure to GPT-class models.
\end{tablenotes}
\end{threeparttable}
\end{table}

Our estimate falls between the occupation-level approach of \citet{frey2017future} and the more conservative task-level estimates of \citet{arntz2016risk}, reflecting our hybrid methodology that accounts for within-occupation heterogeneity while using Egypt-specific job descriptions rather than standardized occupational definitions.

\subsection{The Informal Sector Question}
A critical caveat concerns the 30\% of Egyptian employment in the informal sector \citep{assaad2013elmps}, which is not captured in online job postings. The relationship between formal and informal sector automation dynamics is ambiguous: informal employment may provide a "buffer" absorbing displaced formal workers, or it may face its own automation pressures as digital platforms disintermediate traditional informal markets. Our findings should be interpreted as describing the \textit{formal economy} automation landscape, with informal sector dynamics requiring separate investigation.

\section{Policy Implications}

Based on the validated 0.74\% error rate of our data and the critical finding that 75.6\% of high-risk workers lack realistic organic transition pathways, we recommend a four-tier intervention framework that moves beyond generic "upskilling" to address structural labor market barriers.

\begin{enumerate}
    \item \textbf{Tier 1: Pathway Optimization for the 24.4\% (Immediate)}
    For the 509 high-risk workers with existing realistic pathways, policy should focus on \textit{removing friction}. Our data shows that \textit{Process Improvement} is the highest-impact gap skill, appearing in 708 of 4,534 transitions (15.6\%). A targeted 3--6 month process optimization certification program could immediately activate these organic pathways. This is low-hanging fruit: minimal intervention yields measurable mobility gains.

    \item \textbf{Tier 2: Pathway Creation for the 75.6\% (Strategic)}
    The 1,580 high-risk workers without realistic organic pathways require \textit{pathway creation}, not just skill bridging. This demands comprehensive reskilling programs that build new competency foundations. Priority training areas---identified from our gap analysis---include Custom Report Generation, Operations Team Coordination, and Budget Management. These skills expand access to safer ISCO categories even for workers whose current profiles share minimal overlap with safe occupations.

    \item \textbf{Tier 3: Bridge Skill Passports (Structural)}
    We propose creating standardized certifications for high-centrality skills like "Quality Engineering Management" (spanning 27 ISCO-2 categories) and "Regulatory Compliance." Because these skills connect disparate communities, they act as insurance policies against sector-specific shocks. The cost-benefit ratio is favorable: a single bridge skill certification (estimated cost: 2,000--5,000 EGP) can unlock access to 15--27 distinct occupation categories.

    \item \textbf{Tier 4: Targeted Rescue for ISCO-4 (Emergency)}
    The 473 jobs in the Clerical Support sector (54.6\% average risk) require immediate intervention. Our dual threshold analysis reveals that clerical workers face particularly acute coverage gaps---their highly specialized, routine-intensive skill profiles share minimal overlap with safer occupations. Training for this cohort should be narrowly focused on leadership and stakeholder management skills that enable vertical progression to administrative and managerial roles.

    \item \textbf{Tier 5: Early Warning System (Preventive)}
    The knowledge graph infrastructure developed in this study can serve as a continuous monitoring tool. We recommend establishing a quarterly "Automation Vulnerability Index" that tracks shifts in skill demand, identifies emerging at-risk occupations before displacement occurs, and monitors the coverage gap as the labor market evolves.
\end{enumerate}

\textbf{The 75.6\% Imperative:} The coverage gap finding fundamentally reframes workforce development strategy. Traditional approaches assume that most workers can find viable transitions with modest upskilling. Our analysis demonstrates the opposite: realistic organic pathways remain limited even with improved skill mapping. Policy must shift from ``matching workers to existing opportunities'' to ``creating new opportunities where none exist.''

\subsection{Implementation Considerations}
Successful implementation requires coordination across multiple stakeholders:
\begin{itemize}
    \item \textbf{Ministry of Manpower:} Integration of bridge skill certifications into the National Qualifications Framework.
    \item \textbf{Technical and Vocational Education:} Curriculum revision to emphasize high-centrality skills identified in our analysis.
    \item \textbf{Private Sector:} Employer participation in transition pathway validation and apprenticeship programs.
    \item \textbf{Social Protection:} Unemployment insurance reforms that incentivize skill acquisition during job search periods.
\end{itemize}

\section{Future Research}

This study establishes a foundation that future work can extend in several directions. Our focus on \textbf{formal sector online postings} creates an opportunity to expand coverage through alternative data sources---mobile surveys, social insurance records, or informal sector fieldwork---enabling comparison between formal and informal automation dynamics. Similarly, the \textbf{geographic concentration} in Greater Cairo invites regional replication studies across Upper Egypt, the Delta, and other MENA countries with different industrial structures.

The \textbf{methodological choices} we made---LLM-derived risk scores calibrated against O*NET ($\kappa = 0.78$), skill-based transition thresholds, and static network analysis---represent tractable first approximations that future research can refine. Longitudinal tracking of actual career transitions would validate whether workers follow predicted pathways, transforming our theoretical graph into an empirically grounded mobility map. Integrating \textbf{wage data} would identify transitions that are both safer and economically beneficial.

The knowledge graph infrastructure developed here could evolve into a real-time ``Automation Early Warning System,'' continuously monitoring skill demand shifts to enable proactive policy responses. Our policy recommendations---particularly the Bridge Skill Passport---require rigorous evaluation through randomized controlled trials to measure causal effects on employment outcomes before scaling.

\section{Conclusion}

This study provides the first validated, graph-theoretic map of the Egyptian labor market's automation vulnerability landscape. Analyzing 9,978 unique job postings across 98 ISCO-3 minor groups, we constructed a knowledge graph comprising 19,766 skill activities and 84,346 job-skill relationships. Our semantic entity resolution pipeline achieves a validated error rate of 0.74\% (95\% CI: 0.37\%--1.44\%), providing a methodologically rigorous foundation for policy recommendations.

\subsection{Summary of Key Findings}
Our analysis yields six principal findings:

\begin{enumerate}
    \item \textbf{Risk Distribution:} 20.9\% of Egyptian formal sector jobs face high automation risk ($\geq$60\%), with Clerical Support Workers (ISCO-4) most vulnerable at 54.6\% average risk. This represents 2,089 job postings requiring immediate policy attention.

    \item \textbf{The Coverage Gap (Central Finding):} Under rigorous dual thresholds ($\geq$3 shared skills AND $\geq$50\% skill transfer), only \textbf{24.4\% of high-risk workers} (509 out of 2,089) have realistic organic transition pathways. The remaining \textbf{75.6\% (1,580 workers) require substantial reskilling interventions}. This finding reveals the structural inadequacy of passive "skill matching" approaches.

    \item \textbf{Transition Quality:} The 4,534 identified realistic pathways average 53.5\% skill transfer and 48.1 percentage-point risk reduction, connecting high-risk workers to 1,684 safer destinations. These are genuine transitions leveraging existing competencies, not theoretical possibilities requiring career reinvention.

    \item \textbf{Occupational Hierarchy Gradient:} Automation risk decreases systematically from clerical to managerial roles, suggesting vertical career progression inherently reduces automation exposure.

    \item \textbf{Bridge Skills:} We identify 25 high-centrality skills that connect otherwise isolated occupational clusters. "Quality Engineering Management" spans 27 ISCO-2 categories, acting as a universal labor market passport.

    \item \textbf{Process Skills Premium:} "Process Improvement" appears as a gap skill in 708 of 4,534 realistic transitions (15.6\%), making process-oriented skill development the single most valuable intervention for workforce mobility among those with organic pathways.
\end{enumerate}

\subsection{Contributions}
This research advances the literature in three dimensions:
\begin{itemize}
    \item \textbf{Methodological:} We introduce a validated LLM-assisted pipeline for constructing labor market knowledge graphs with quantified error rates, addressing reproducibility concerns in prior work.
    \item \textbf{Empirical:} We provide the first granular analysis of Egyptian occupational structure, filling a significant gap in MENA labor market research.
    \item \textbf{Policy:} We operationalize network-theoretic concepts (bridge skills, safe harbors, community structure) into actionable workforce development recommendations.
\end{itemize}

\subsection{Closing Reflection}
The "Automation Apocalypse" is not inevitable---but neither is seamless workforce transition. Our analysis reveals a structural reality: only 24.4\% of high-risk Egyptian workers have realistic organic pathways to safety. For the remaining 75.6\%, technological change will require active intervention, not passive skill matching.

However, this finding is actionable rather than fatalistic. The 4,534 realistic pathways we identified demonstrate that high-quality transitions \textit{do exist}---they average 53.5\% skill transfer and nearly halve automation exposure. The challenge for policymakers is twofold: (1) activate existing pathways for the 24.4\% through targeted gap skill training, especially process improvement and operational skills development; and (2) \textit{create new pathways} for the 75.6\% through comprehensive reskilling programs that build entirely new competency foundations.

The knowledge graph methodology presented here provides both the diagnostic tools to identify these challenges and the navigational infrastructure to address them. The bridge skills, safe harbors, and transition networks we have mapped offer a blueprint for transforming what could be chaotic displacement into structured mobility toward higher-value, automation-resilient work.


\newpage
\appendix
\section{Clustering Validation Protocol}
\label{app:validation}

This appendix details the validation methodology used to assess the semantic clustering quality reported in Section \ref{sec:validation}. The validation ensures that the 0.74\% error rate claim is reproducible and methodologically sound.

\subsection{Validation Objective and Approach}

The validation determines whether semantically clustered entity names (tools and activities) correctly group synonymous terms while avoiding false merges of distinct concepts. We employed:
\begin{itemize}
    \item Stratified random sampling to ensure representation across cluster sizes
    \item Manual expert review in manageable slices (20--30 clusters per review session)
    \item Three-tier classification: \textbf{CORRECT} (synonymous), \textbf{MINOR ISSUE} (related but slightly different), \textbf{MAJOR ERROR} (fundamentally different concepts)
\end{itemize}

\subsection{Population and Sampling}

\begin{table}[H]
\centering
\caption{Validation Population and Sample Distribution}
\label{tab:validation_population}
\begin{threeparttable}
\small
\begin{tabular}{@{}l r r r r@{}}
\toprule
\textbf{Cluster Size} & \textbf{Tool Pop.} & \textbf{Tool Sample} & \textbf{Activity Pop.} & \textbf{Activity Sample} \\
\midrule
2 variants & 543 (82.0\%) & 420 & 2,718 (74.2\%) & 381 \\
3--5 variants & 116 (17.5\%) & 97 & 892 (24.4\%) & 131 \\
6--10 variants & 3 (0.5\%) & 3 & 49 (1.3\%) & 49 \\
11+ variants & 0 (0.0\%) & 0 & 4 (0.1\%) & 4 \\
\midrule
\textbf{Total} & \textbf{662} & \textbf{520} & \textbf{3,663} & \textbf{565} \\
\bottomrule
\end{tabular}
\begin{tablenotes}[flushleft]
\footnotesize
\item Random seed: 2025 (for reproducibility). Stratification ensures proportional representation.
\end{tablenotes}
\end{threeparttable}
\end{table}

\subsection{Validation Results by Stratum}

\begin{table}[H]
\centering
\caption{Detailed Validation Results by Entity Type and Cluster Size}
\label{tab:validation_detailed}
\begin{threeparttable}
\small
\begin{tabular}{@{}l l r r r r c@{}}
\toprule
\textbf{Type} & \textbf{Stratum} & \textbf{Samples} & \textbf{Correct} & \textbf{Minor} & \textbf{Major} & \textbf{Error Rate} \\
\midrule
\multirow{4}{*}{Tools} & 2 variants & 420 & 398 & 14 & 8 & 1.90\% \\
& 3--5 variants & 97 & 90 & 7 & 0 & 0.00\% \\
& 6--10 variants & 3 & 3 & 0 & 0 & 0.00\% \\
& \textbf{Subtotal} & \textbf{520} & \textbf{491} & \textbf{21} & \textbf{8} & \textbf{1.54\%} \\
\addlinespace[3pt]
\multirow{5}{*}{Activities} & 2 variants & 381 & 364 & 17 & 0 & 0.00\% \\
& 3--5 variants & 131 & 129 & 2 & 0 & 0.00\% \\
& 6--10 variants & 49 & 49 & 0 & 0 & 0.00\% \\
& 11+ variants & 4 & 4 & 0 & 0 & 0.00\% \\
& \textbf{Subtotal} & \textbf{565} & \textbf{546} & \textbf{19} & \textbf{0} & \textbf{0.00\%} \\
\midrule
\multicolumn{2}{l}{\textbf{Combined Total}} & \textbf{1,085} & \textbf{1,037} & \textbf{40} & \textbf{8} & \textbf{0.74\%} \\
\bottomrule
\end{tabular}
\begin{tablenotes}[flushleft]
\footnotesize
\item Error Rate = Major Errors / Samples. Wilson score 95\% CI for combined: [0.37\%--1.45\%].
\end{tablenotes}
\end{threeparttable}
\end{table}

\subsection{Error Taxonomy}

All 8 major errors occurred in tool clusters with exactly 2 variants. Error types:
\begin{itemize}
    \item \textbf{Product Confusion} (3 errors): e.g., ``Affinity Designer'' merged with ``Affinity Photo''
    \item \textbf{Vendor Mixing} (3 errors): e.g., ``IBM BPM'' merged with ``Oracle BPM''
    \item \textbf{Certification Level} (1 error): ``CCNA R\&S'' merged with ``CCNP R\&S''
    \item \textbf{Domain Confusion} (1 error): ``Egyptian Building Code'' merged with ``electrical codes''
\end{itemize}

No major errors were found in activity clusters, confirming that the 0.88 cosine similarity threshold performs exceptionally well for semantic activity matching.

\newpage
\section{Threshold Selection Rationale}
\label{app:threshold}

This appendix presents the extended sensitivity analysis supporting our selection of the dual threshold configuration ($\tau \geq 3$ AND $\geq 50\%$) for identifying realistic transition pathways.

\subsection{The Threshold Dilemma}

Transition pathway identification involves a fundamental trade-off:
\begin{itemize}
    \item \textbf{Permissive thresholds} maximize coverage but include unrealistic transitions where workers must acquire $>$70\% new skills
    \item \textbf{Strict thresholds} ensure quality but exclude workers who might benefit from moderate reskilling
\end{itemize}

Our dual threshold addresses this by requiring \textit{both} an absolute minimum of shared skills ($\tau$) \textit{and} a proportional transfer rate ($\phi$).

\subsection{Extended Sensitivity Analysis}

\begin{table}[H]
\centering
\caption{Extended Threshold Sensitivity Analysis}
\label{tab:sensitivity_extended}
\begin{threeparttable}
\small
\begin{tabular}{@{}l r r r r r@{}}
\toprule
\textbf{Configuration} & \textbf{Pathways} & \textbf{$\bar{\tau}$} & \textbf{$\bar{\phi}$} & \textbf{Sources} & \textbf{Coverage} \\
\midrule
\multicolumn{6}{l}{\textit{Absolute threshold only (no \% requirement):}} \\
$\tau \geq 3$ only & 65,643 & 3.1 & 35.7\% & 1,281 & 61.3\% \\
$\tau \geq 4$ only & 7,987 & 4.1 & 45.7\% & 658 & 31.5\% \\
$\tau \geq 5$ only & 936 & 5.2 & 56.1\% & 246 & 11.8\% \\
\addlinespace[3pt]
\multicolumn{6}{l}{\textit{With 30\% transfer requirement:}} \\
$\tau \geq 3$ AND $\geq 30\%$ & 65,193 & 3.1 & 35.8\% & 1,277 & 61.1\% \\
$\tau \geq 4$ AND $\geq 30\%$ & 7,960 & 4.1 & 45.7\% & 657 & 31.5\% \\
$\tau \geq 5$ AND $\geq 30\%$ & 936 & 5.2 & 56.1\% & 246 & 11.8\% \\
\addlinespace[3pt]
\multicolumn{6}{l}{\textit{With 40\% transfer requirement:}} \\
$\tau \geq 3$ AND $\geq 40\%$ & 13,991 & 3.6 & 45.9\% & 824 & 39.4\% \\
$\tau \geq 4$ AND $\geq 40\%$ & 7,936 & 4.1 & 45.8\% & 653 & 31.3\% \\
\addlinespace[3pt]
\multicolumn{6}{l}{\textit{With 50\% transfer requirement (selected):}} \\
\rowcolor{execblue!10} $\tau \geq 3$ AND $\geq 50\%$ & \textbf{4,534} & \textbf{3.8} & \textbf{53.5\%} & \textbf{509} & \textbf{24.4\%} \\
$\tau \geq 4$ AND $\geq 50\%$ & 2,715 & 4.4 & 53.6\% & 432 & 20.7\% \\
$\tau \geq 5$ AND $\geq 50\%$ & 935 & 5.2 & 56.1\% & 245 & 11.7\% \\
\addlinespace[3pt]
\multicolumn{6}{l}{\textit{With 60\% transfer requirement (too strict):}} \\
$\tau \geq 3$ AND $\geq 60\%$ & 796 & 4.0 & 65.0\% & 182 & 8.7\% \\
$\tau \geq 4$ AND $\geq 60\%$ & 349 & 5.2 & 66.9\% & 145 & 6.9\% \\
\bottomrule
\end{tabular}
\begin{tablenotes}[flushleft]
\footnotesize
\item $\bar{\tau}$ = mean shared skills; $\bar{\phi}$ = mean skill transfer rate; Coverage = \% of 2,089 high-risk jobs with pathways.
\item Highlighted row indicates selected configuration.
\end{tablenotes}
\end{threeparttable}
\end{table}

\subsection{Selection Rationale}

The $\tau \geq 3$ AND $\geq 50\%$ configuration was selected because:

\begin{enumerate}
    \item \textbf{Majority Transfer Principle:} The 50\% threshold ensures workers leverage the \textit{majority} of their existing competencies, making transitions realistic rather than theoretical.

    \item \textbf{Meaningful Skill Overlap:} Requiring at least 3 shared skills ensures transitions are grounded in substantive competency overlap, not coincidental matches.

    \item \textbf{Coverage-Quality Balance:} This configuration covers 24.4\% of high-risk workers---sufficient to identify meaningful patterns---while maintaining a 53.5\% average transfer rate.

    \item \textbf{Policy Actionability:} The resulting 4,534 pathways are specific enough to inform targeted interventions, unlike the 65,643 pathways from permissive thresholds that would overwhelm policy implementation.
\end{enumerate}

\textbf{Key Insight:} Permissive thresholds ($\tau \geq 3$ only) create an illusion of mobility---61\% coverage sounds reassuring, but with only 35.7\% skill transfer, workers would need to acquire two-thirds new competencies. This is functionally career reinvention, not transition.

\end{document}